\documentclass[manuscript]{aastex}

\newcommand{\mha}{MH$\alpha$}
\newcommand{\Lkha}{LkH$\alpha$}

\usepackage{url}
\usepackage{multirow}

\makeatletter
\newcommand{\rmnum}[1]{\romannumeral #1}
\newcommand{\Rmnum}[1]{\expandafter\@slowromancap\romannumeral #1@}
\makeatother

\begin{document}

\title{A FAR-ULTRAVIOLET ATLAS OF LOW-RESOLUTION {\it HST} SPECTRA OF T TAURI STARS\altaffilmark{1}}

\altaffiltext{1}{Based on observations made with the NASA/ESA \emph{Hubble Space Telescope}, obtained from the data
    archive at the Space Telescope Science Institute. STScI is operated by the Association of Universities for
    Research in Astronomy, Inc. under NASA contract NAS 5-26555. This work also contains results from Chandra projects 
    09200763 and 10200804 supported by SAO grants GO8-9024X and GO9-0020B to the University of Colorado. 
    This work was supported in part by NASA Swift grants NNX09AL59G and NNX10AK88G, and Smithsonian 
    Institution Chandra grants GO8-9024X, GO9-0020B, GO0-11042X, and GO1-12031X to the University of Colorado.}


\author{Hao Yang}
\affil{JILA, University of Colorado and NIST, Boulder, CO 80309-0440} 
\email{haoyang@jilau1.colorado.edu}

\author{Gregory J. Herczeg}
\affil{Max-Planck-Institut f\"ur extraterrestriche Physik, Postfach 1312, 85741 Garching, Germany} 
\email{gregoryh@mpe.mpg.de}

\author{Jeffrey L. Linsky}
\affil{JILA, University of Colorado and NIST, Boulder, CO 80309-0440} 
\email{jlinsky@jilau1.colorado.edu}

\author{Alexander Brown}
\affil{CASA, University of Colorado, Boulder, CO 80309-0389} 
\email{Alexander.Brown@colorado.edu}

\author{Christopher M. Johns-Krull}
\affil{Department of Physics and Astronomy, Rice University, 6100 Main Street, Houston, TX 77005} 
\email{cmj@rice.edu}

\author{Laura Ingleby}
\affil{Department of Astronomy, University of Michigan, 830 Dennison Building, 500 Church Street, Ann Arbor, MI 48109} 
\email{lingleby@umich.edu}

\author{Nuria Calvet}
\affil{Department of Astronomy, University of Michigan, 830 Dennison Building, 500 Church Street, Ann Arbor, MI 48109} 
\email{ncalvet@umich.edu}

\author{Edwin Bergin}
\affil{Department of Astronomy, University of Michigan, 830 Dennison Building, 500 Church Street, Ann Arbor, MI 48109} 
\email{ebergin@umich.edu}

\begin{abstract}

We present a far-ultraviolet (FUV) spectral atlas consisting of spectra of 91 pre-main sequence stars. 
Most stars in this sample were observed with the Space Telescope Imaging Spectrograph (STIS) and Advanced Camera for Surveys (ACS) 
on the \emph{Hubble Space Telescope} (\emph{HST}). A few archival spectra from \emph{International Ultraviolet Explorer} 
(\emph{IUE}) and the Goddard High Resolution Spectrograph (GHRS) on the \emph{HST} are included for completeness. 
We find strong correlations among the \ion{O}{1} $\lambda$1304 triplet, 
the \ion{Si}{4} $\lambda\lambda$1394/1403 doublet, the \ion{C}{4} $\lambda$1549 doublet, and the \ion{He}{2} $\lambda$1640 line luminosities.
For classical T Tauri stars (CTTSs), we also find strong correlations between these lines and the accretion luminosity, 
suggesting that these lines form in processes related to accretion.
These FUV line fluxes and X-ray luminosity correlate loosely with large scatters. 
The FUV emission also correlates well with H$\alpha$, H$\beta$, and \ion{Ca}{2} K line luminosities.
These correlations between FUV and optical diagostics can be used to obtain rough estimates of FUV line fluxes from optical observations.
Molecular hydrogen (H$_{2}$) emission is generally present in the spectra of actively accreting CTTSs 
but not the weak-lined T Tauri stars (WTTSs) that are not accreting. 
The presence of H$_2$ emission in the spectrum of HD 98800 N suggests that the disk should be classified as actively accreting rather than a debris disk.  
We discuss the importance of FUV radiation, including the hydrogen Ly$\alpha$ line, on the photoevaporation of exoplanet atmospheres.
We find that the \ion{Ca}{2}/\ion{C}{4} flux ratios for more evolved stars are lower than those for less evolved accretors, 
indicating preferential depletion of refactory metals into dust grains. 

\end{abstract}


\section{INTRODUCTION}

Classical T Tauri stars (CTTSs) are young, pre-main sequence stars characterized by excess line and continuum emission 
produced by a circumstellar disk and accretion from the disk onto the central star. Their non-accreting counterparts, 
weak-lined (or naked) T Tauri stars (WTTSs or NTTSs), generally do not display excess infrared (IR) emission, indicating the absence of a dusty disk.
The spectra of WTTSs also lack strong line and excess hydrogen continuum emission, indicating that no accretion is present.
In recent years, large photometric and spectroscopic surveys of T Tauri stars (TTSs) at wavelengths spanning from X-rays through the millimeter range, 
utilizing many line and continuum diagnostics, have demonstrated that 
(\rmnum{1}) most young stars lose their disks within a few million years \citep[e.g.][]{Haisch2001,Andrews2005,Hernandez2008}, 
(\rmnum{2}) stars that do not show excess mid-IR emission, which indicates the presence of warm dust, 
also lack accretion signatures \citep{White2001,Muzerolle2003,Fedele2010}, 
(\rmnum{3}) accretion processes are similar for central objects with masses ranging 
from a solar mass down to brown-dwarf masses \citep{Muzerolle2005,Mohanty2005}, 
(\rmnum{4}) the Initial Mass Function peaks at $\sim0.5$ $M_\odot$ \citep{Luhman2003,Luhman2010}, 
and (\rmnum{5}) all CTTSs and WTTSs are coronally active \citep[e.g., ][]{Feigelson2005,Guedel2007}.
These surveys, often focusing on stars within the nearby Taurus Molecular Cloud, have provided the foundation for understanding the early evolution 
of our own and other planetary systems.

Although TTSs have been surveyed at most accessible wavelengths, the characterization of far-ultraviolet (FUV) emission remains sparse.
FUV spectra of young stars offer intriguing diagnostics of accretion, magnetic activity, outflows, and 
disks \citep[e.g.,][]{cmj2000,Herczeg2002,Lamzin2004,cmjgreg2007}. The emission produced by these processes plays a particularly important role 
in the evolution of circumstellar disks where planetary systems originate. FUV emission produced by stellar chromospheric\footnotemark[1]~activity
and accretion shocks causes the disk to photoevaporate at large distances during the accretion phase, removing gas in the outer disks and
perhaps constricting the time a disk survives \citep[e.g.,][]{Gorti2009} and thus the time available for planet formation. FUV emission also 
modulates the chemistry at the disk surface by dissociating H$_2$O, CO, and HCN molecules and ionizing some species (C, Si, S) with low-ionization 
potentials \citep{Aikawa2002,Bergin2003,Bergin2004,Bethell2009}.  Some important gas tracers, such as the prominent [\ion{O}{1}] $63 \mu$m line, 
have fluxes that depend directly on the FUV luminosity of the central star \citep[e.g.,][]{Woitke2009,Kamp2011}.  
The ionization at the disk surface may play a role in allowing the disk to accrete via the  magneto-rotationally instability \citep{PerezBecker2011}.
Once accretion has ceased, FUV emission is produced only by stellar magnetic activity. Measuring FUV emission also provides important constraints 
on the amount of EUV emission produced by young stars, which controls the survival timescale for any small amount of gas in a remnant 
disk \citep{Alexander2005,Alexander2006}.  After a disk has dissipated and planets have formed, the EUV and FUV emission causes some 
evaporation of the atmospheres of ``hot Jupiters'' \citep{Lecavelierdesetangs2003,Linsky2010} and remains an important input for atmospheric
chemistry in planetary atmospheres \citep[e.g.,][]{Yelle2004}.
\footnotetext[1]{For convenience, in the paper we refer both chromospheric emission lines (e.g., \ion{Ca}{2} and Balmer lines)
and lines emitted from the transition region (e.g., \ion{C}{4} and \ion{Si}{4} lines) as chromospheric.}

The \emph{International Ultraviolet Explorer} (\emph{IUE}) pioneered UV spectroscopy of pre-main sequence (PMS) stars, revealing bright 
emission in many lines, including \ion{He}{2} $\lambda1640$, \ion{C}{4} $\lambda1549$, \ion{C}{2} $\lambda1335$, \ion{O}{1} $\lambda1304$, 
and the rich fluorescent spectrum of molecular hydrogen (H$_{2}$) \citep{brown1981}. \citet{Valenti2000}, \citet{cmj2000}, and \citet{Valenti2003} published 
the PMS archive of {\it IUE} far- and near-UV spectra of 137 TTSs and 97 Herbig Ae/Be (HAeBe) stars, 
although only 50 of the TTSs were observed in the FUV. Together, 
this trilogy has laid out the foundation for our understanding of UV radiation fields from young stars, including the strength and the physical processes responsible for such emission.

However,  the {\it IUE} survey of FUV emission from young stars was limited to the FUV-brightest CTTSs and included only three WTTSs, 
of which the latest spectral type is K0. The survey is also limited by S/N, evident in the low detection rate of H$_2$ line emission \citep[13/32 CTTSs, ][]{Valenti2000} despite subsequent observations showing that such emission is common to all 
CTTSs \citep{Herczeg2006,Ingleby2009}.
Analysis of the location and kinematics of the gas that produces the FUV emission lines was also limited by the large 
aperture ($\sim 10\arcsec \times 20\arcsec$) and low spectral resolution (6 \AA) of the {\it IUE} SWP camera.

Since the launch of the \emph{Hubble Space Telescope} (\emph{HST}) 20 years ago, the Goddard High Resolution Spectrograph (GHRS),
Space Telescope Imaging Spectrograph (STIS) and the Advanded Camera for Surveys (ACS) prisms have 
observed over 80 PMS stars. Analyses of small subsets of these observations ($<10$ objects, and often only one) have been used to address specific issues, 
including (\rmnum{1}) the details of Ly$\alpha$-pumped H$_2$ emission \citep[e.g.][]{Ardila2002,Herczeg2002, Herczeg2004,Herczeg2006}, 
(\rmnum{2}) the H$_2$ emission resulting from collisions with energetic electrons \citep{Bergin2004,Herczeg2004,Ingleby2009,France2010b}, 
(\rmnum{3}) the influence of Ly$\alpha$ emission and X-ray emission on disk chemistry \citep{Bergin2003,Bergin2004,Bethell2009}, 
(\rmnum{4}) the ionization state of outflows \citep{cmjgreg2007}, 
(\rmnum{5}) the possibility of metal depletion in the accretion flow \citep{Herczeg2002,France2010},
and (\rmnum{6}) the origin of emission from ionized gas \citep[e.g.][]{Ardila2002,Herczeg2002,Lamzin2004}.  
In a more complete analysis of $\sim40$ CTTSs and WTTSs observed with the ACS PR130L prism, \citet{Ingleby2009} placed very low limits 
on the amount of remnant H$_2$ gas around WTTSs by finding that H$_2$ continuum emission is always detected from CTTSs but never from WTTSs, 
including those that retain debris disks but are no longer accreting.  Recently, \citet{Ingleby2011} analyzed ten new ACS/SBC spectra 
of Chamaeleon I and II regions along with archival HST data of $\sim$45 TTSs 
and showed that the FUV emission decreases with age, correlating with the decline of accretion in CTTSs as they become nonaccretors. 
Of the many FUV observations of CTTSs and WTTSs obtained by {\it HST}, only a small subsample, those obtained with the E140M grating of 
STIS ($R \sim 40,000$), have also been published in a spectral atlas, the CoolCat catalog 
\footnote{\url{http://casa.colorado.edu/~ayres/CoolCAT}} \citep{Ayres2005}.

In this paper, we present an atlas of FUV spectra and emission line fluxes of all PMS stars observed 
by STIS, GHRS and ACS. This atlas surveys the strength of FUV emission from WTTSs and includes CTTSs 
with a wide range of accretion rates and masses. Since {\it HST} will not be able to obtain large amounts of FUV observations to 
correlate with extensive ground-based observations, we search for correlations between the currently available FUV observations and optical 
observations of the same sample to serve as a good guide for future studies. 
With these correlations, large volumes of optical data sets \citep[e.g.,][]{Barentsen2011} can be used
to infer the FUV emission of more distant or heavily extincted TTSs.
In \S\ 2, we describe sample selection and the details of FUV observations with various instruments.
In \S\ 3, we present flux measurements of strong atomic features, the correlation between the atomic line luminosities and stellar properties, 
and identification of molecular hydrogen features. In \S\ 4, we discuss estimating FUV emission from optical observations, evolution of FUV emission
from PMS stars, importance of FUV emission on the photoevaporation of exoplanet atmospheres, and the interesting special case of HD 98800 N.
A summary of our findings is provided in \S\ 5.


\section{SAMPLE SELECTION AND OBSERVATIONS}
Table \ref{obslog} lists the 91 PMS stars in our sample and information concerning their observations.
The majority of the observations were obtained with the STIS G140L grating (24 stars) and ACS/SBC PR130L 
prism (54 stars). The observations also include eight stars observed with the STIS E140M grating 
and two stars observed only with the GHRS G160M grating. 
Three stars, CY Tau (ID: 9), DR Tau (ID: 38), and GM Aur (ID: 40) were observed with both the ACS/SBC PR130L prism and the STIS E140M grating.
T Tau (ID: 14) was observed with both the STIS G140L and E140M gratings.
For completeness, six stars that were observed with \emph{IUE} but not with {\it HST} are included in our sample.
Among the six stars, MML 34 (ID:73) was not published in the \citet{Valenti2000} \emph{IUE} atlas of PMS stars.

\subsection{Sample Description}
Table~\ref{pmstable} summarizes the basic properties of stars in our sample.
The sample consists of a collection of CTTSs\footnotemark[1] and WTTSs that have been observed 
with {\it HST} in the FUV. The high-resolution STIS spectra and the IUE spectra are biased 
toward stars that are brightest in the UV. Most of the stars are members of the Taurus Molecular Cloud.  
A small number are members of other regions, such as the TW Hya association (TWA), or are isolated. 
(A few tight binaries in the sample are further discussed in Appendix \ref{binary}.)
The spectral types in the sample range from F2 to M8, which corresponds to masses 
of $\sim$ 2 $M_\odot$ down to $\sim$ 0.1 $M_\odot$.  Accretion rates of the objects range from $10^{-12}$ to $10^{-7}$ $M_\odot$ yr$^{-1}$.
\footnotetext[1]{For the purposes of this paper, the terms CTTSs and WTTSs are used to apply to objects with both stellar and brown dwarf masses.}

All members of a molecular cloud are assumed to have the same distance: $140$ pc 
for Taurus \citep{Bertout1999,Loinard2007}, $125$ pc for Ophiuchus \citep{Lombardi2008,Loinard2008}, $145$ pc 
for Upper Sco \citep{dezeeuw1999}, $130$ pc for Corona Australus \citep{Neuhauser2008}, $150$ pc 
for Lupus I \citep{Comeron2009}, $175$ pc for Cha I \citep[see discussion in][]{Luhman2008}, 
$250$ pc for Perseus \citep{Enoch2006}, $440$ pc for Ori OB1c,  and $450$ pc for $\lambda$ Ori \citep{dolan2001}.  
TW Hya (ID: 57) has a {\it Hipparcos} distance of $56$ pc from \citet{Perryman1997}, and the brown dwarf 2M1207A (ID: 69) has a 
parallax distance of $52.4$ pc from \citet{Ducourant2007}. The remaining TW Hya association (TWA) members have kinematic distances 
calculated by \citet{Mamajek2005}.

In most cases, the spectral types and photospheric and accretion luminosities were adopted from the references listed in Table~\ref{pmstable}, 
with some changes to account for updated distance measurements. For cases where the accretion rate is obtained from  
\citet{White2001}, we convert the accretion rate into the accretion luminosity using the listed accretion rate, stellar radius, and 
stellar mass and the formula $\dot{M}_{acc}=  1.25~L_{acc}R_* / (GM_*)$ from \citet{Gullbring1998}.  
In several cases, e.g., DR Tau (ID: 38), DP Tau (ID: 36) and DL Tau (ID: 26), 
the photospheric luminosity is not listed because the detected optical/IR emission is dominated by emission from the 
accretion shocks and the disk. These objects are frequently listed as ``continuum'' objects in the literature, though some 
have spectral types that have been obtained by measuring heavily veiled photospheric absorption lines in high-resolution optical spectra.

Extinctions are obtained from the listed references in Table~\ref{pmstable}.  The uncertainty in extinction estimates is often not listed.  
For the purposes of this paper, we assume that most extinctions are uncertain by $\sim 0.5$ mag, which introduces 
a factor of $\sim 6$, $\sim 4$, and $\sim 2$ error in luminosities of the \ion{O}{1}  $\lambda1304$, \ion{C}{4} $\lambda1549$, 
and \ion{Ca}{2} K lines, respectively.  In some cases, such as members of the TWA, the extinction is negligible 
and does not contribute any uncertainty to the line luminosities.

For a few stars in the sample, two or four exposures were taken consecutively by STIS, as indicated by the 
third-to-last column (N $=$ number of exposures) in Table~\ref{obslog}.  We examined the individual spectra and did not find
substantial short-term variability between successive exposures. We therefore coadded the spectra, weighted by exposure time, and
then measured the line fluxes. For AU Mic, there are $\sim$ 180 spectra available in the archive, and we picked 32 spectra from 
two consecutive days and coadded them.

For the four stars observed with both STIS G140L and ACS or both STIS G140L and E140M gratings, 
we used the FUV luminosities from the STIS G140L spectra for correlations,
because of its better balance between spectral resolution and signal-to-noise.
The total FUV luminosities measured from {\it IUE} are typically unreliable for CTTSs and WTTSs 
because the noise is high relative to the continuum and the detector often saturates at longer wavelengths.

\subsection{Observations with Different Instruments}

Table~\ref{instruments} summarizes the properties of the different instruments that were used to obtain the 
data for this survey, including aperture size, wavelength coverage, spectral resolution $R ~(= \lambda/\delta\lambda$) and 
typical \ion{C}{4} flux levels below which the measurement uncertainty becomes greater than 10\% (see discussion in \S 4.1 and Appendix~\ref{instrument}).
Below, we describe briefly the characteristics of each instrument as well as its data access and reduction procedures.


\subsubsection{HST GHRS}
GHRS was a first-generation {\it HST} FUV spectrograph. The GHRS observations of CTTSs were obtained 
at moderate resolution through the large ($\sim 2\arcsec$) aperture
and covered only $\sim 35$~\AA\ regions, which were centered on the bright \ion{C}{4} doublet at 1550 \AA\ and the bright 
\ion{Si}{4} doublet at 1400 \AA. The GHRS spectra of eight CTTSs were published by \citet{Ardila2002}.
 We obtained these spectra from the Multimission Archive at the Space Telescope Science Institute 
(MAST; \url{http://archive.stsci.edu}). The spectra were reduced using the standard On The Fly Reprocessing (OTFR) 
pipeline at MAST \citep{swade2001} before being downloaded from the archive.

\subsubsection{HST STIS E140M}
The second-generation {\it HST} instrument, STIS, is an echelle spectrograph that offers high spectral resolution ($ R = \lambda/\delta\lambda \sim 40,000$) 
across the 1170--1700 \AA\ wavelength region in the E140M mode. When observing the young stars, the echelle mode of STIS used 
very small apertures ($0.2\arcsec \times 0.06\arcsec$, $0.2\arcsec \times 0.2\arcsec$, and $0.5\arcsec \times 0.5\arcsec$), which 
minimizes any contribution from spatially extended emission. STIS E140M spectra of six CTTSs were published by \citet{Herczeg2006}. 
The eight young stars observed with STIS E140M grating in our sample were obtained from the CoolCAT spectral atlas of cool stars \citep{Ayres2005}.

\subsubsection{HST STIS G140L}  
The G140L grating on STIS is a low-resolution ($R\sim2000$) spectrograph that covers the 1150--1700 \AA\ wavelength region. 
A total of 24 CTTSs were observed with STIS/G140L, typically with the $52\arcsec \times2\arcsec$ aperture. The spectra 
were reduced using the standard OTFR pipeline before being downloaded from the MAST archive.
Most STIS G140L spectra in our sample were published by \citet{Calvet2004} and \citet{Bergin2004}.

\subsubsection{HST ACS/SBC PR130L}
The Solar Blind Channel (SBC) on ACS observed 54 TTSs with the PR130L prism, which yields a spectral resolution of  
$ R \sim 170$ near 1350 \AA\ but decreasing to $ R \sim 60$ at 1650 \AA. Each visit consists of an image with one of several long-pass filters 
and a prism spectrum. The 2D images were obtained from the MAST archive, as calibrated by the OTFR pipeline. We extracted the spectra
using custom IDL routines, with a wavelength solution and extraction window based on the positional 
offsets between the location of the object in the image and the slitless spectrum \citep{larsen2006-2}.  
The counts spectrum was obtained using a 9-pixel extraction window centered on the spectral trace and was then converted 
to a flux spectrum based on the sensitivity function calculated by \citet{larsen2006-2}. Of the 54 ACS/SBC PR130L spectra, 30 spectra
were published by \citet{Ingleby2009}.

\subsubsection{FUSE and IUE Spectra of TW Hya and AU Mic}
The Far Ultraviolet Spectroscopic Explorer ({\it FUSE}) covers the 912--1184 \AA\ region with $R\sim 20,000$ 
and a $30^{\prime\prime}\times 30^{\prime\prime}$  aperture \citep{Moos2000}. We present fluxes from {\it FUSE} 
spectra of TW Hya (ID: 57), a CTTS, and AU Mic (ID: 90), a WTTS, to provide readers with an estimate 
of the total FUV flux down to 912 \AA\ for both sources.  {\it FUSE} observed AU Mic on 26 August 2000 and 
10 October 2001 for a total of 43.8 ks \citep{Redfield2003,France2007}.  
{\it FUSE} observed TW Hya on 3 June 2000 and again, with a much deeper spectrum, on 20-21 February 2003 \citep{cmjgreg2007}.  

We present the \emph{IUE} LWR-LO spectra of TW Hya and AU Mic to show typical near-UV fluxes of TTSs from 1700 to 3000 \AA. 
\emph{IUE} observed TW Hya on 29 October 1979 for 3600 s. \emph{IUE} observed AU Mic at multiple epochs. We averaged the spectra
from $18$ exposures on 4--6 August 1980 for a total of 32.4 ks.

\subsubsection{Optical Spectra}
We supplemented the HST/FUV observations with ground-based optical spectra of many of the same targets, obtained with the Double Spectrograph \citep{Oke1982} on the Hale 5m Telescope on Mount Palomar in January and December 2008. The optical  spectra span most of the 3200--8700 \AA\ range with $R\sim1000$. Fluxes in H$\alpha$, H$\beta$, and \ion{Ca}{2} H lines were extracted for this paper, with an uncertainty in flux calibration of 10\%. The full dataset will be described in a future paper.

\subsubsection{X-Ray Luminosity}

In Table 2, we include the available measurements of X-ray luminosity for the UV sample obtained from both the literature and 
new dedicated observations. Many of the measurements for stars in the Taurus-Auriga star formation region are taken from the XMM-Newton XEST survey 
\citep{Guedel2007}.  

New Chandra ACIS observations of LkCa 4 (ID: 8), DE Tau (ID: 12), GM Aur (ID: 40), TWA 13 N and S (ID: 61 and 62), TWA 8 S and N (ID: 65 and 66 ), 
and TWA 9 N and S (ID: 67 and 68) and Swift observations of TWA 7 (ID: 55) and TWA 3 S and N (ID: 58 and 59 ) were analyzed for 
this paper and the derived X-ray luminosities are included in Table 2.  The Chandra and Swift data were modeled using XSPEC (Version 12.5) and 
fluxes and luminosities over the energy range 0.3-10.0 keV were measured.

\section{Results}

\begin{figure}[ht]
  \begin{center}
    \includegraphics[scale=0.75]{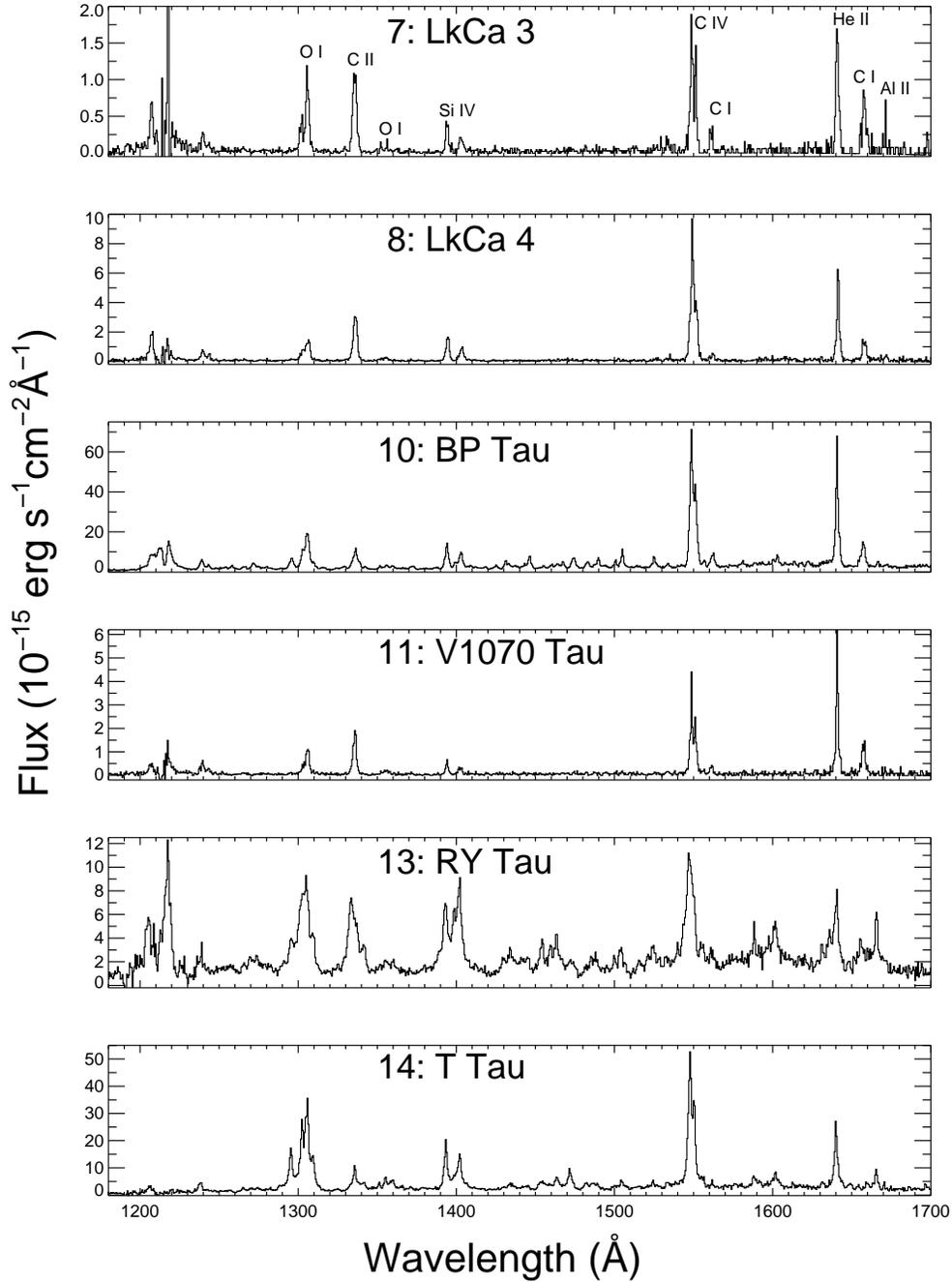}
     \caption{ The STIS G140L spectra for a few stars in our sample are shown. Plots of the full sample are available in the online version of this paper.}
           \label{stisplot1}
              \end{center}
    \end{figure}

\begin{figure}[ht]
  \begin{center}
    \includegraphics[scale=0.65]{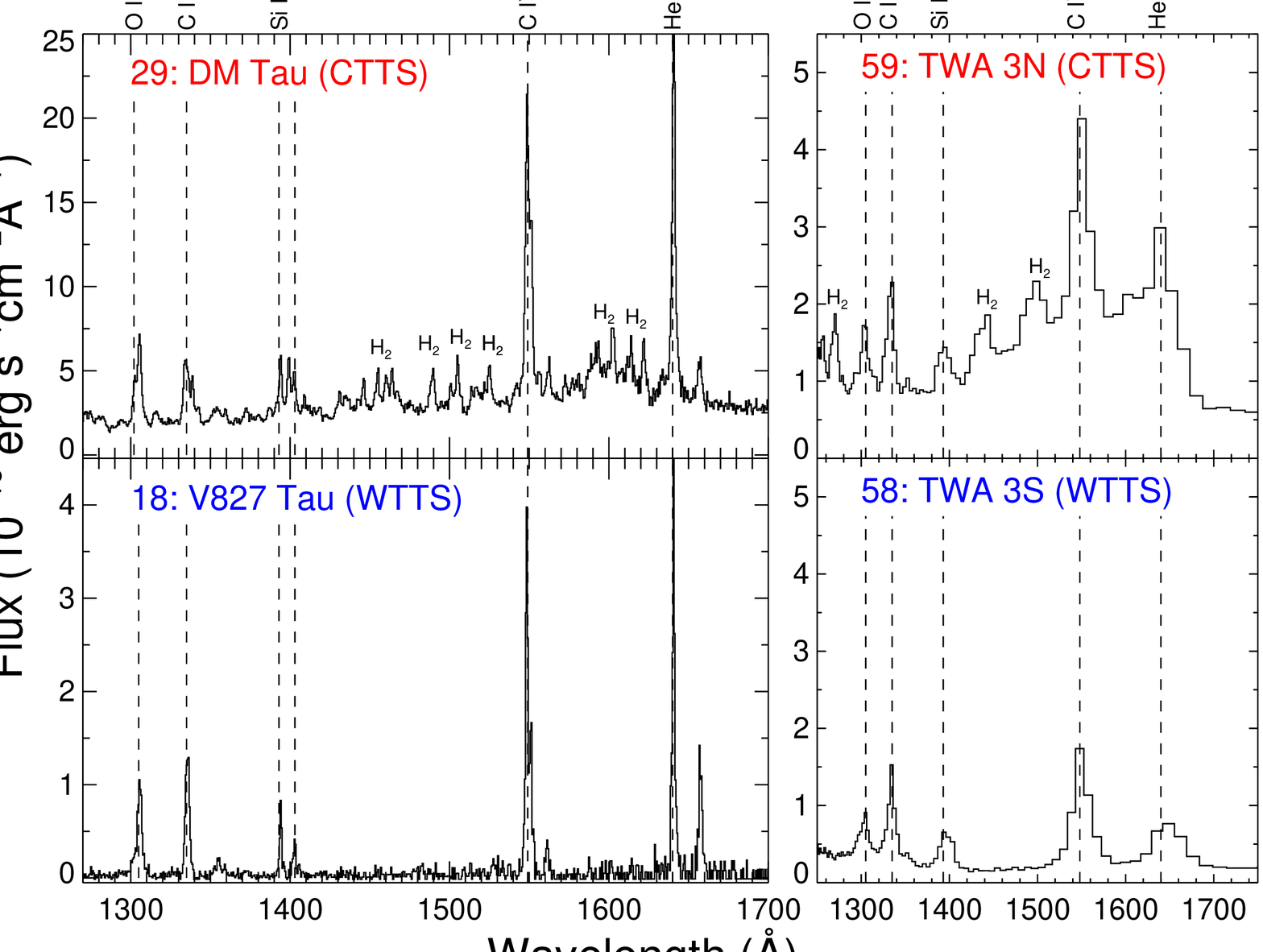}
       \caption{ STIS G140L spectra of DM Tau and V827 Tau are shown in the left panels, and ACS PR130L spectra of TWA 3N and TWA 3S
              are shown in the right panels. Prominent atomic and molecular emission features are marked.
                Plots of the full sample are available in the online version of this paper. }
           \label{samplespec}
              \end{center}
    \end{figure}

Figure \ref{stisplot1} shows the STIS G140L spectra for a few stars in our sample, and plots of the full sample are available in the online 
version of this paper.
Figure \ref{samplespec} compares typical FUV spectra of CTTSs and WTTSs. The pattern of FUV emission from the WTTSs is similar to that of magnetically active 
dwarf stars, including bright emission in \ion{O}{1} $\lambda1304$, \ion{C}{2} $\lambda1335$, \ion{Si}{4} $\lambda1400$, \ion{C}{4} $\lambda1549$, 
and \ion{He}{2} $\lambda1640$ lines and no detectable continuum emission. (The ACS spectra of WTTSs appear to show continuum, but that is due to
a combination of low spectral resolution and strong line emission and is not real .)
The CTTSs show strong emission in these same lines, though the origin 
is likely an accretion-heated photosphere or funnel flow \citep{Calvet1998} rather than chromospheric emission. The CTTSs also show bright H$_{2}$
emission lines and, in many cases, continuum emission \citep{France2011}.

In the following subsections, we measure flux in emission lines and discuss several exceptional cases where at least some of the FUV emission is 
spatially extended in outflows rather than on or very near the star.

\subsection{Line Flux Measurements}
We measured line fluxes by subtracting the continuum estimated with a linear fit with wavelength, and subsequently integrating the emission 
remaining in the line. The uncertainties in our measurements are dominated by uncertainty in the placement of the continuum, with the error determined 
by measuring new fluxes after increasing and decreasing the estimated continuum by 20\%. This arbitrary change in continuum estimate is an conservative 
estimate and makes only a minor contribution 
to the uncertainty in line luminosities, which are usually dominated by uncertainties in the FUV extinction.
Statistical errors propagated through the data reduction pipelines are also included.
For previously published flux measurements, 
we adopt the values from \citet{Valenti2000}, \citet{Pagano2000}, \citet{Ardila2002}, and \citet{Calvet2004}.
Table~\ref{allflux} lists the line flux measurements and associated uncertainties for the stars in our sample. 


Accurately measuring line fluxes in the very low-resolution ACS/SBC PR130L spectra is complicated by line-blending and 
uncertain continuum subtraction. We created artificial ACS spectra by degrading 21 different STIS G140L spectra to the resolution of PR130L. 
Line fluxes were then measured in both the real and the degraded G140L spectra, and we show the differences between the two sets of 
measurements in Table~\ref{pseudoacs}. 
The line fluxes have differences of $<10$\% in most cases but can reach as high as $30$\%, especially for the \ion{C}{4} $\lambda1549$ doublet 
where the ACS/SBC PR130L spectral resolution is lowest.

\begin{figure}[ht]
  \begin{center}
    \includegraphics[scale=0.75]{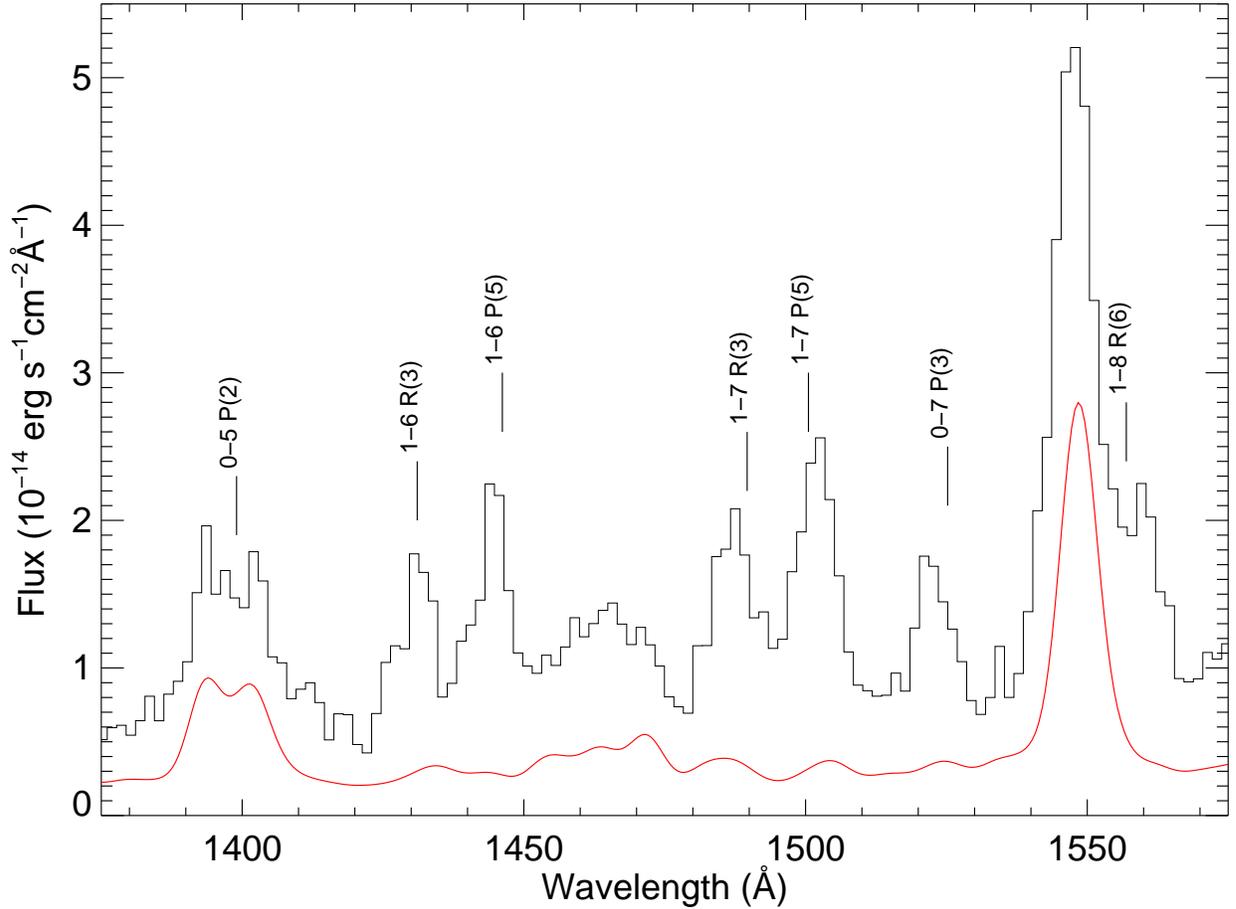}
       \caption{ FUV spectra of T Tau observed with two different instruments and aperture sizes.
             The histogram is the observation with the \emph{IUE} SWP-LO grating using the large ($10\arcsec\times20\arcsec$) aperture, and
            the solid red curve is a STIS G140L spectrum observed with a $52\arcsec\times0.2\arcsec$ slit and degraded to the \emph{IUE} spectral resolution. 
             Several notable H$_2$ features in the \emph{IUE} spectrum are identified with solid vertical lines, 
             and the transitions are also noted.}
           \label{aperture}
              \end{center}
    \end{figure}

The FUV emission is typically consistent with a point source centered on the star, with some notable exceptions.
Figure~\ref{aperture} compares the spectra of T Tau (ID: 14) obtained with a narrow, on-source extraction from the STIS G140L grating (degraded
to the IUE spectral resolution) and with the IUE spectrum, obtained with a large aperture.  
The {\it IUE} spectrum shows much brighter H$_2$ emission than the on-source G140L spectrum.  
Bright H$_2$ emission is spatially extended in the G140L slit, consistent with emission produced in the outflow 
(see detailed analysis in \citet{Walter2003} and \citet{Saucedo2003}).   DP Tau (ID: 36) was faintly detected 
in the F165LP acquisition image.  Strong H$_2$ emission is located approximately at the stellar 
position and is spatially extended by $\sim 0\farcs11$ in the cross-dispersion direction (-30$^\circ$) and $\sim 0\farcs14$ in the dispersion direction.  Some faint emission may also be more extended in the spectral image.
The ACS SBC acquisition image for 2MASS J04141188+2811535 (ID: 3), an accreting M6.5 brown dwarf, was obtained with the F140LP long-pass filter, 
which covers a spectral region that includes \ion{C}{4} and H$_2$ lines and the continuum.  
A second source is located 0\farcs34 from the primary at a PA of $\sim 176^\circ$ with a brightness three times fainter than the primary.

\subsection{Correlation of the Line Luminosities with Stellar Properties}

Correlations of line luminosities with stellar properties can be used to infer the physical origin of the emission.  
\citet{cmj2000} found for a small sample of 12 CTTSs a correlation between mass accretion rate  and \ion{C}{4} emission and ascribed the emission to 
accretion-related processes. 
In addition, \citet{cmj2000} and \citet{Ardila2002} suggested also from small samples that FUV emission from WTTSs is weaker than that from CTTSs.  
With a sample of 56 stars, \citet{Ingleby2011} found that both the \ion{C}{4} and the \ion{He}{2} line luminosities correlate well with accretion luminosity,
supporting the connection between the line formation and accretion.
In this subsection, we look for correlations between FUV line emission and available optical chromospheric activity indicators:
\ion{Ca}{2} K, H$\alpha$, and H$\beta$ lines.

We convert the FUV line fluxes to line luminosities using the distances listed in Table~\ref{pmstable}, after correcting the line fluxes
for interstellar extinction following the \citet{cardelli1989} extinction law with a total-to-selective extinction coefficient of $R_{V}$ of $3.09$.
The uncertainty in $A_{V}$ and the extinction law contribute significant uncertainties to the line luminosities. The extinction law for dense molecular clouds 
typically has $R_{V}\sim5.5$, indicating significant grain growth \citep[e.g.][]{Weingartner2001,Indebetouw2005}.  
However, these large values of $R_{V}$ typically correspond to $A_{V}>10$ (also see discussion in \citet{Calvet2004}), 
while most stars in our sample have $A_{V}<2$.  
Some lines-of-sight may also pass through a disk atmosphere, which would likely have different grains and a different extinction curve than the ISM.
Although we choose $R_{V}=3.09$, selecting other values would be reasonable. A full assessment of different extinction laws 
would require recalculating $A_{V}$ and is beyond the scope of this work.
The extinction-corrected FUV luminosity between 1250--1700 \AA\ is listed in the last column of Table~\ref{allflux}.  

\begin{figure}[ht]
  \begin{center}
    \includegraphics[scale=0.65]{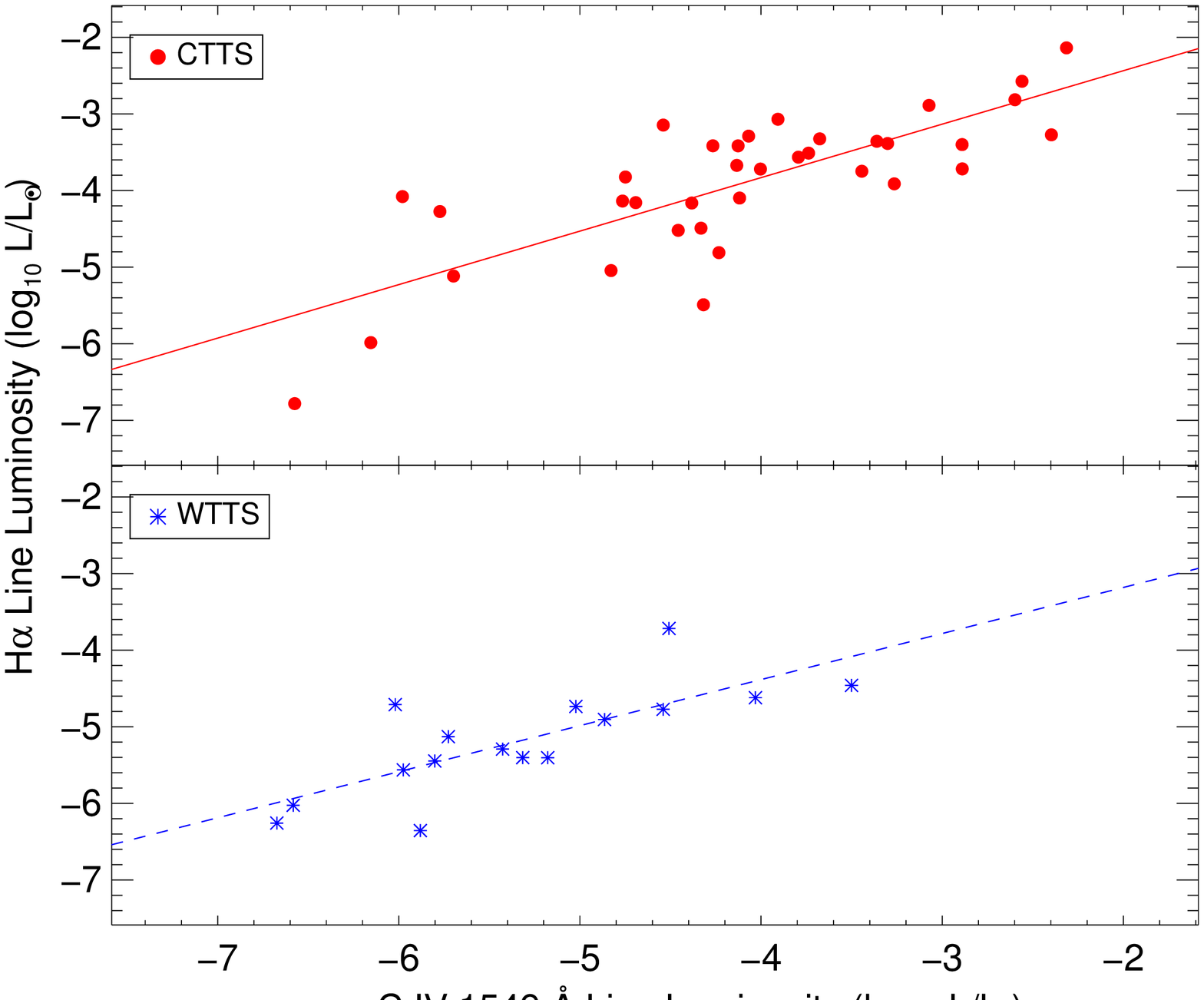}
       \caption{ H$\alpha$ line luminosity is plotted against the \ion{C}{4} $\lambda$1549 doublet luminosity for CTTSs 
              (\emph{top panel}) and WTTSs (\emph{bottom panel}). The red solid and blue dashed lines are the best fits in logarithmic space
               for CTTSs and WTTSs, respectively.}
           \label{civhalpha}
              \end{center}
    \end{figure}

\begin{figure}[ht]
  \begin{center}
    \includegraphics[scale=0.65]{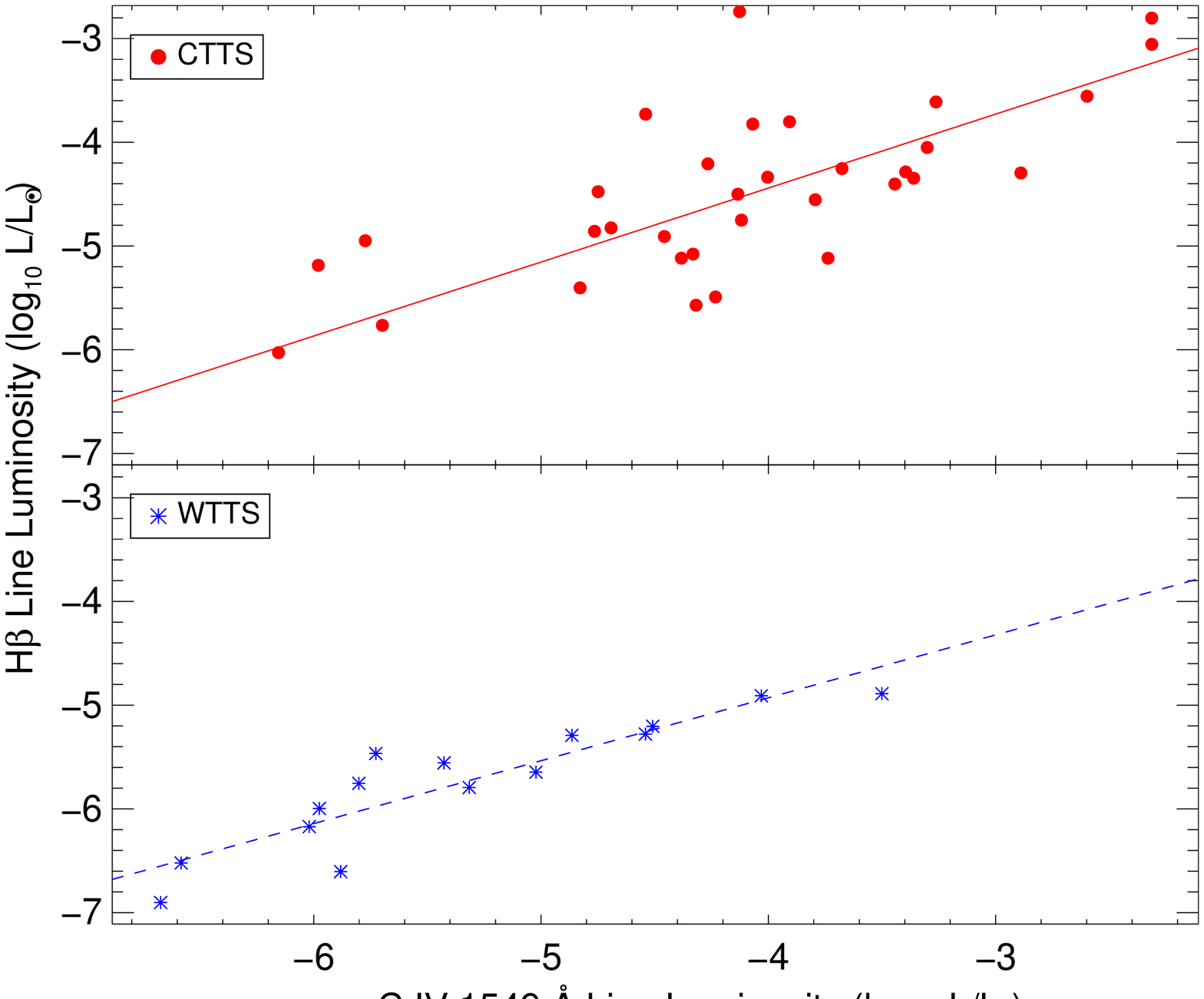}
       \caption{ H$\beta$ line luminosity is plotted against the \ion{C}{4} $\lambda$1549 doublet luminosity for CTTSs 
              (\emph{top panel}) and WTTSs (\emph{bottom panel}). The red solid and blue dashed lines are the best fits in logarithmic space
               for CTTSs and WTTSs, respectively.}
           \label{civhbeta}
              \end{center}
    \end{figure}

\begin{figure}[ht]
  \begin{center}
    \includegraphics[scale=0.65]{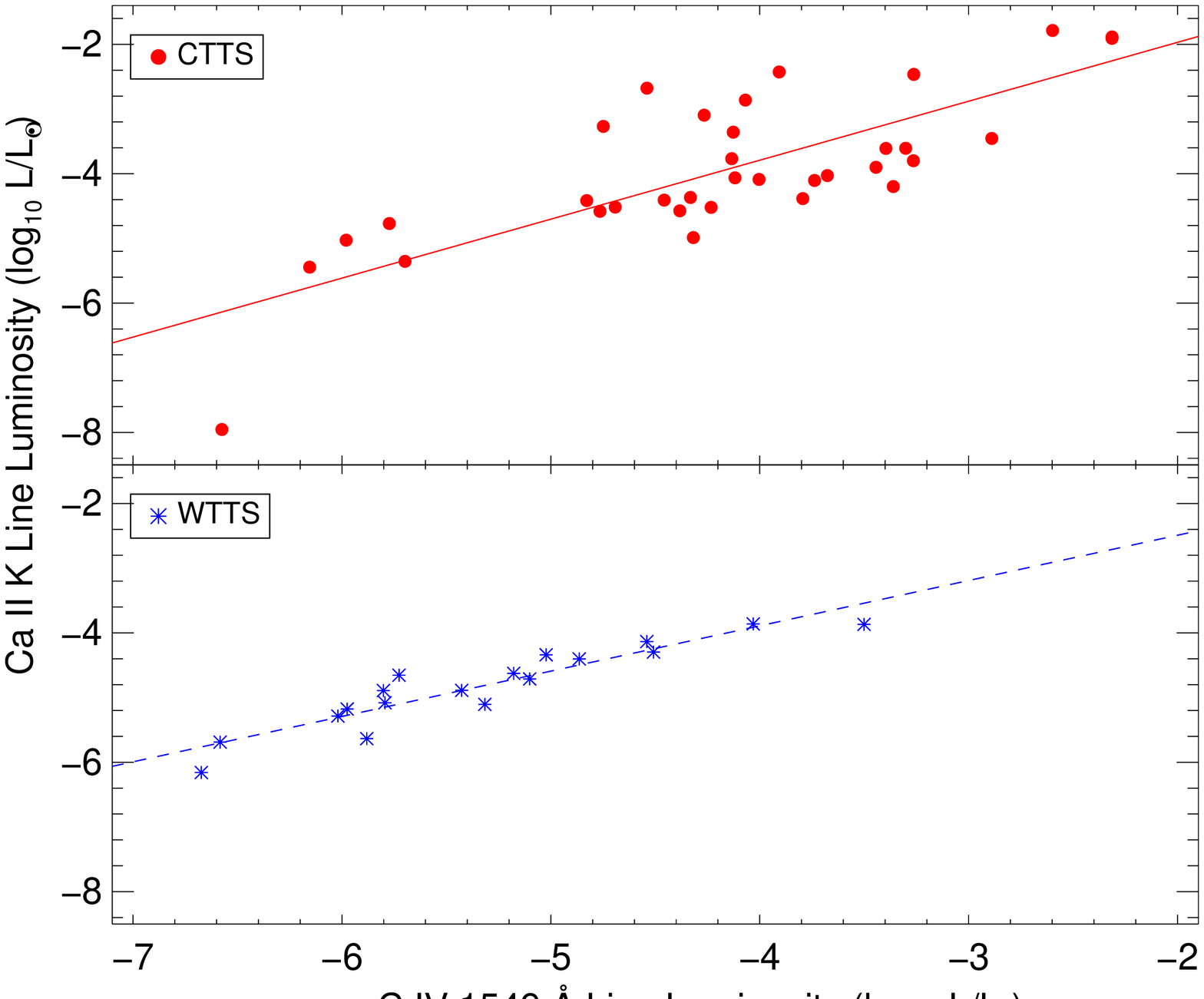}
       \caption{ The \ion{Ca}{2} K line luminosity is plotted against the \ion{C}{4} $\lambda$1549 doublet luminosity for CTTSs 
              (\emph{top panel}) and WTTSs (\emph{bottom panel}). The red solid and blue dashed lines are the best fit in logarithmic space
               for CTTSs and WTTSs, respectively.}
           \label{civcaiik}
              \end{center}
    \end{figure}

\begin{figure}[ht]
  \begin{center}
    \includegraphics[scale=0.65]{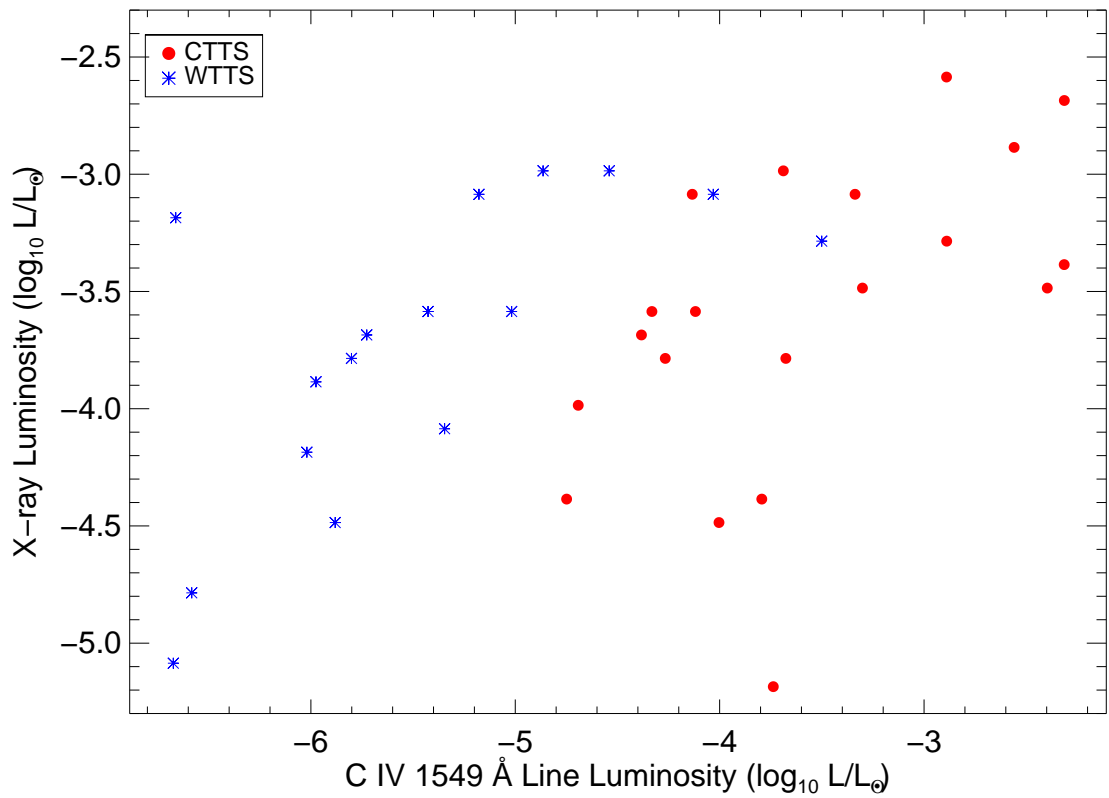}
       \caption{The X-ray luminosity is plotted against the \ion{C}{4} $\lambda$1549 doublet luminosity.
           CTTSs are plotted as red filled circles and WTTSs are plotted as blue asterisks.}
           \label{xrayciv}
              \end{center}
    \end{figure}

Figures~\ref{civhalpha}, \ref{civhbeta}, and \ref{civcaiik} show the \ion{C}{4} $\lambda$1549 line luminosity against H$\alpha$, H$\beta$,
and \ion{Ca}{2} K line luminosities, respectively. For WTTSs, the \ion{C}{4} $\lambda$1549 luminosities are well correlated 
with the optical line luminosities.  For CTTSs, the \ion{C}{4} $\lambda$1549 and the optical line luminosities are loosely correlated, 
which is expected from non-simultaneous observations because the accretion rate varies significantly with time.
The best linear-fit parameters and median scatters from the fits are listed in Table~\ref{linefits} where 
we also list the parameters for linear fits with a fixed slope of one.  
We find that the X-ray emission is loosely correlated with the \ion{C}{4} $\lambda1549$ emission, as shown in Figure~\ref{xrayciv}. 
Variability in both the FUV and X-ray emission from CTTSs should introduce a large scatter in any correlation.  
While the CTTSs show much stronger \ion{C}{4} emission than the WTTSs, mainly due to accretion, 
the CTTSs and the WTTSs display a similar range of X-ray emission.

\begin{figure}[ht]
  \begin{center}
    \includegraphics[scale=0.65]{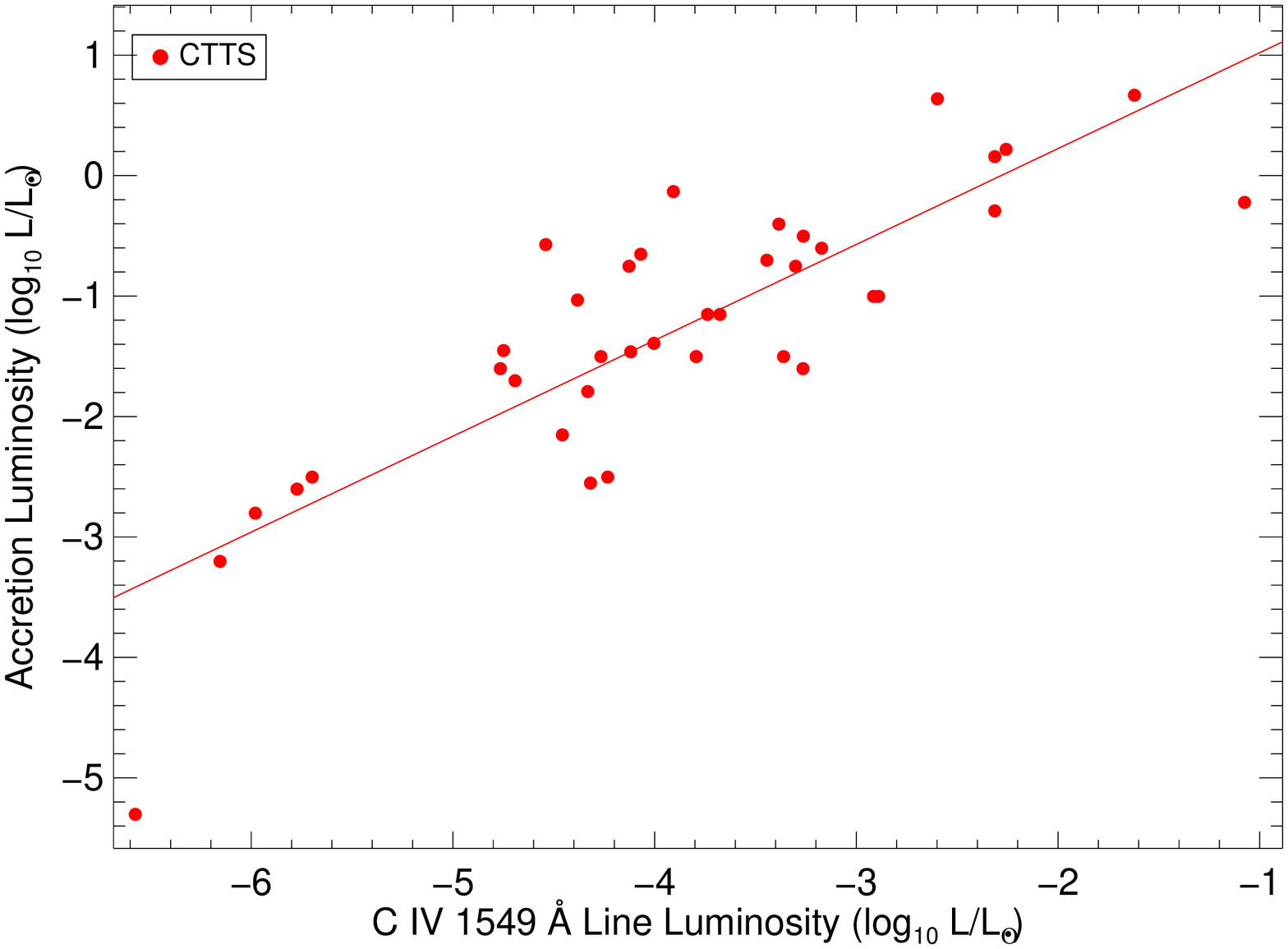}
       \caption{ The accretion luminosity is plotted against the \ion{C}{4} $\lambda$1549 doublet luminosity for CTTSs.
               The red solid line is the best fit in logarithmic space.}
           \label{accretionciv}
              \end{center}
    \end{figure}

\begin{figure}[ht]
  \begin{center}
    \includegraphics[scale=0.65]{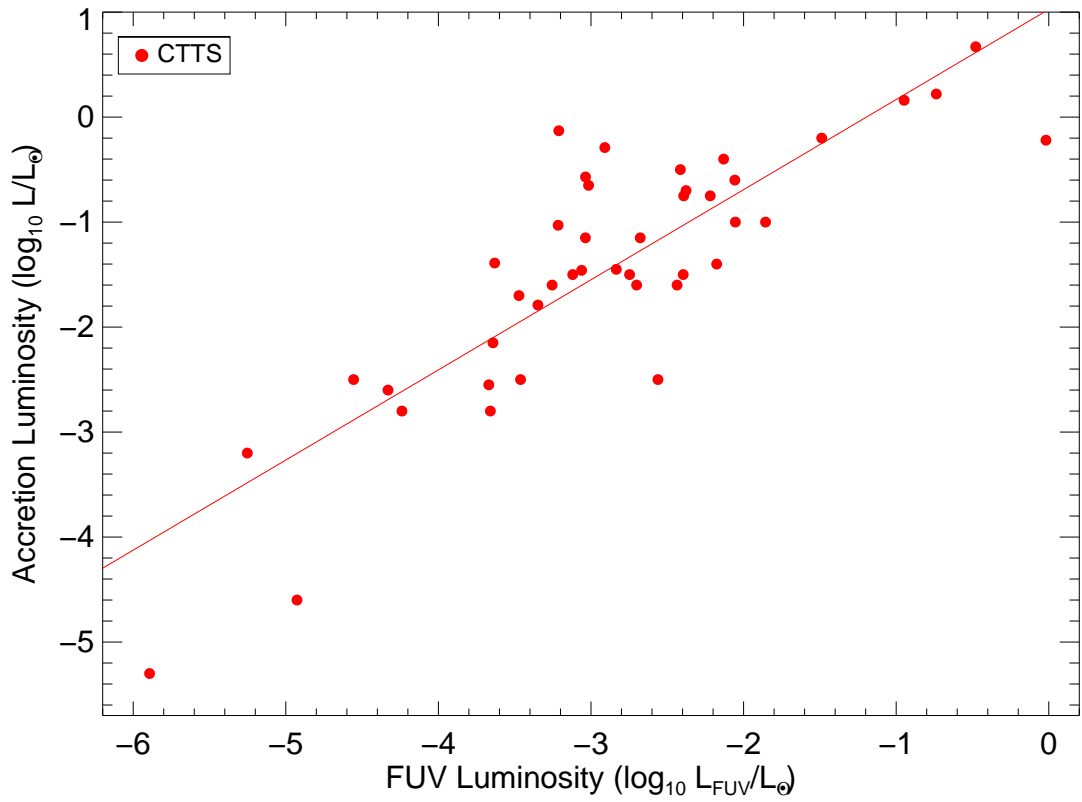}
       \caption{ The accretion luminosity is plotted against the FUV luminosity for CTTSs.
               The red solid line is the best fit in logarithmic space.}
           \label{accretionlfuv}
              \end{center}
    \end{figure}

Figures~\ref{accretionciv} and \ref{accretionlfuv} show correlations of \ion{C}{4} $\lambda$1549 line luminosity and total FUV luminosity
with the accretion luminosity for 37 CTTSs, respectively. 
The \ion{C}{4} emission of CTTSs includes contributions from chromospheric activity and accretion, both of which can be variable. 
The large scatter in the fits likely results from variability between the \ion{C}{4} and accretion rate measurements obtained at different times, 
uncertainty in extinction, and perhaps different physical processes contributing to the \ion{C}{4} line flux. Nonetheless, the correlations between
\ion{C}{4} and both the accretion and FUV luminosities are definitely real. 

Figure~\ref{linecorrelation} shows correlations among various FUV emission line luminosities for both CTTSs and WTTSs.  
The same linear fit applies to both CTTSs and WTTSs, which demonstrates that FUV line flux ratios do not depend on whether the emission 
is produced by chromospheric activity or by accretion.  The line fluxes depend on the emission measure versus temperature.  
The same linear fit to the CTTSs and WTTSs in each panel indicate that the shape of the emission measure distribution 
must be similar for both chromospheric activity and accretion.

\subsection{Molecular Hydrogen Emission from CTTSs}
Warm H$_2$ gas ($\sim 2,500$ K) can be excited to the B and C electronic states by strong Ly$\alpha$ and other UV emission lines and subsequently 
decays to the ground (X) electronic state. This H$_2$ fluorescence is commonly detected from CTTSs, likely originating 
at the surfaces of the circumstellar disks and from outflows shocking the surrounding molecular environment \citep[e.g.,][]{Walter2003,Saucedo2003,Herczeg2006}.  
The spectral window between 1430 \AA\ and 1510 \AA\ is particularly useful because it includes many strong H$_2$ lines but few stellar emission lines.  
Limited by the low spectral resolution in ACS data, we could only identify the H$_2$ features in the $1430$--$1510$ \AA\ wavelength 
region for the stars in our sample, and noted ``Yes'' or ``No'' in the 
second-to-last column of Table \ref{allflux} to indicate whether or not the spectrum shows H$_2$ emission.  
A more quantitative assessment and analysis of H$_2$ emission from ACS/SBC spectra was provided by \citet{Ingleby2009}.

Our results agree with \citet{Ingleby2009} who find that CTTSs generally display H$_2$ emission in the the $1430$--$1510$ \AA\ region while WTTSs do not.
Of the 31 stars in our sample previously classified as WTTS, only one (HD 98800 N, ID: 63) shows the H$_2$ emission feature (see \S 4.2).  For 
the near-equal mass binary system 2MASS J16141107-2305362 \citep{Metchev2009}, the S component (ID: 83) produces stronger FUV emission lines 
but lacks detectable H$_2$ emission, characteristic of WTTSs.  The N component (ID: 84), produces strong Ly$\alpha$-pumped H$_2$ 
lines and excess FUV continuum emission, as expected for a CTTS.  The system may therefore be a CTTS/WTTS binary.

\section{DISCUSSION}

\subsection{Prescriptions for Calculating FUV luminosities}

FUV emission affects the disk by photoevaporation and by driving a complex chemistry at the disk surface, including photodissocation of some molecules 
and ionization of atomic species with low first ionization potential energies.  The FUV radiation field is needed as an input for disk modeling to help interpret emission line tracers of gas in the disk.

\begin{figure}[ht]
  \begin{center}
    \includegraphics[scale=0.75]{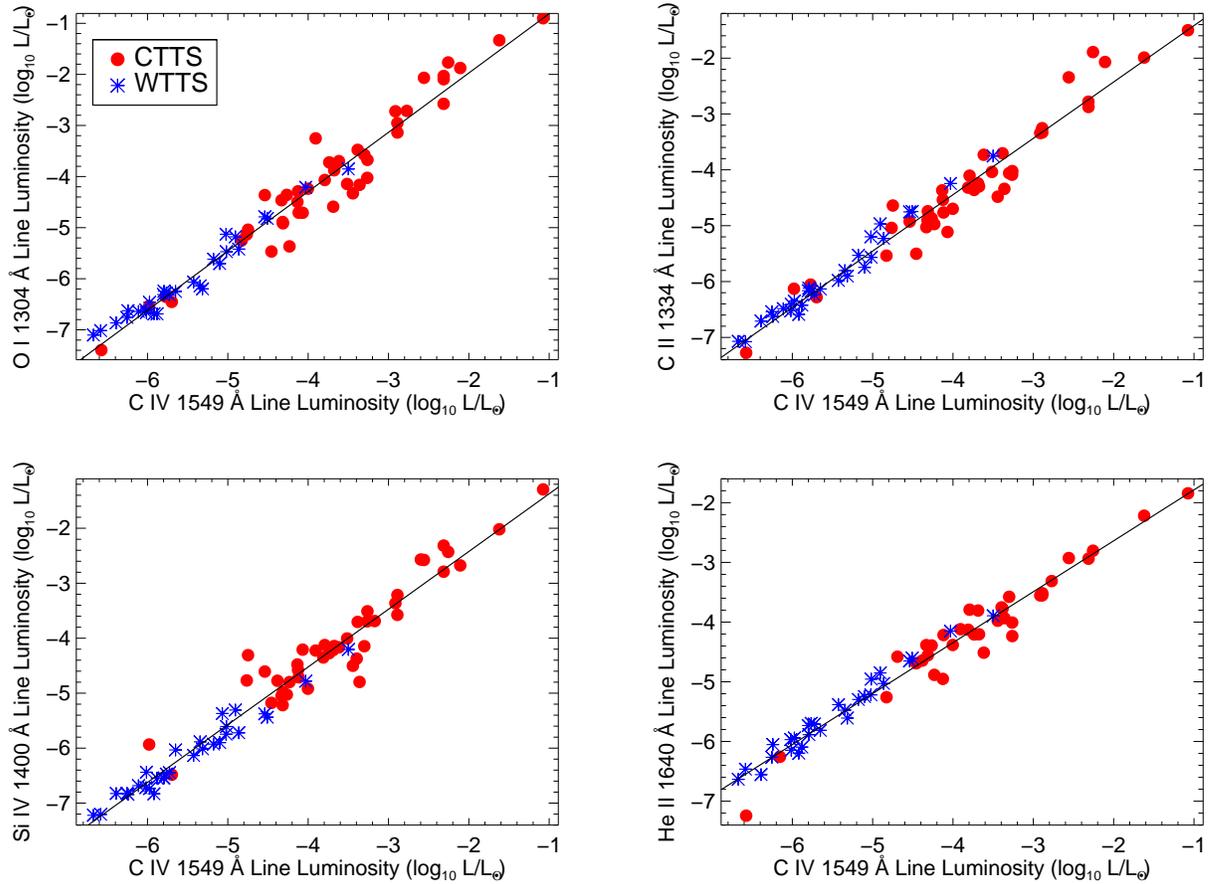}
       \caption{ The \ion{C}{4} $\lambda$1549 doublet luminosity is plotted against the \ion{O}{1} $\lambda$1304 multiplet,
            \ion{C}{2} $\lambda$1335 doublet, \ion{Si}{4} $\lambda$1400 doublet, and \ion{He}{2} $\lambda$1640 line luminosities.
           CTTSs are plotted as red filled circles and WTTSs are plotted as blue asterisks.
            The solid line in each panel is the best fit to all of the data in logarithmic space.}
           \label{linecorrelation}
              \end{center}
    \end{figure}

\begin{figure}[ht]
  \begin{center}
    \includegraphics[scale=0.65]{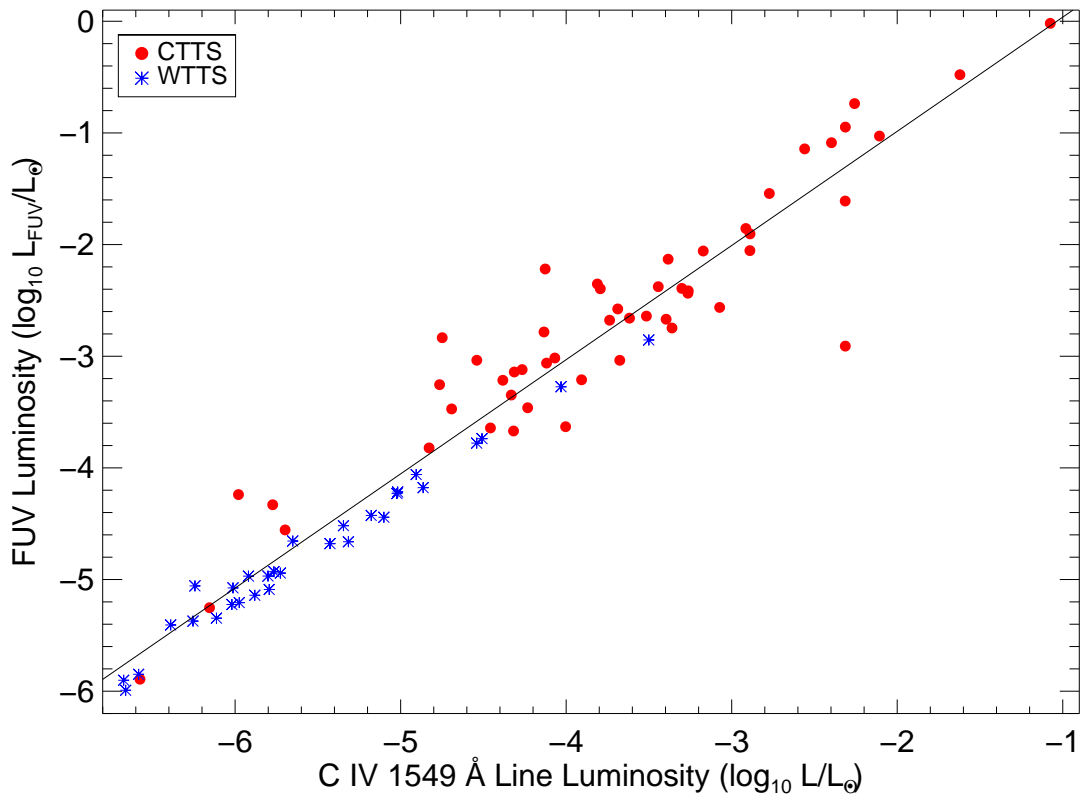}
       \caption{ FUV luminosity is plotted against \ion{C}{4} $\lambda$1549 doublet luminosity.
           CTTSs are plotted as red filled circles. WTTSs are plotted as blue asterisks. 
           The solid line is a best fit to all of the data in logarithmic space.}
           \label{civfuvlum}
              \end{center}
    \end{figure}

In this subsection, we offer prescriptions for estimating accretion and chromospheric emission produced in the FUV from optical emission lines. 
Figures~\ref{linecorrelation} and \ref{civfuvlum} show that the luminosities of \ion{C}{4} $\lambda$1549, \ion{O}{1} $\lambda$1304, \ion{C}{2} $\lambda$1335,
\ion{Si}{4} $\lambda\lambda$1394/1403 lines, and \ion{He}{2} $\lambda$1640 lines and of the FUV (1250--1700 \AA\ bandpass) 
are all strongly correlated with each other. As a consequence, 
a measurement of only one of these parameters is needed to estimate the total FUV flux. Figures~\ref{civhalpha}, \ref{civhbeta}, and \ref{civcaiik} show that 
the \ion{C}{4} and, by implication, the FUV luminosities correlate with H$\alpha$, H$\beta$, and \ion{Ca}{2} K line emission for both CTTSs and WTTSs.  
Emission in H$\alpha$, H$\beta$, and \ion{Ca}{2} K lines can therefore be used to obtain rough estimates for the FUV emission using 
equations and parameters listed in Table~\ref{linefits}.

Two exemplary cases, TW Hya (CTTS) and AU Mic, are selected for this analysis, because they have complete, high S/N spectra from the Lyman limit 
to 1700 \AA\ (see Table~\ref{twhyaaumic}) and the photospheric emission is negligible in this bandpass for both stars.  
The spectra from 912--1200 \AA\ were observed by {\it FUSE} for these two stars but not for most of the stars presented in this atlas.  
However, because the correlation between FUV lines is very tight (Figure~\ref{linecorrelation}), we infer that these two template spectra 
can be scaled to the strength of \ion{C}{4} (or optical line) flux for each object.

%

For example, when a measurement of accretion luminosity is available, the equation
$log(L_{\rm{FUV}})={-1.670}+{0.836}~log(L_{{acc}})$ can be used
to obtain an estimate of the total FUV luminosity. Similarly, the equations $log(L_{\rm{FUV}})={-1.632}+{0.760}~log(L_{\rm{H_{\alpha}}})$ and 
$log(L_{\rm{FUV}})={-2.323}+{0.766}~log(L_{\rm{H_{\beta}}})$ provide estimates of total FUV luminosity from H$_{\alpha}$ and H$_{\beta}$ measurements of
CTTSs, respectively. In Table~\ref{linefits}, we give the relationships of optical diagnositics,  $L_{acc}$, 
and other FUV lines, with respect to \ion{C}{4} emission or $L_{\rm{FUV}}$.

This prescription outlined above provides estimates of the FUV luminosity, with an accuracy of $\sim 0.5-1.0$ dex.
The luminosities refer only to chromospheric and accretion-related processes and do not include photospheric emission.  
For stars with spectral types G and earlier, the photospheric emission provides a significant fraction of the total FUV luminosity 
and needs to be included in such calculations.  Since the FUV spectra of late-type stars are dominated by line emission, calculating excitation rates in discrete electronic transitions of various molecules, including H$_2$ and CO, may require use of real spectral templates rather than assumptions about a smooth distribution of flux with wavelength.

This discussion of FUV radiation fields neglects the Ly$\alpha$ $\lambda$1216 line.  The Ly$\alpha$ line is presumed to be strong, as much as 75-90\% 
of the total FUV luminosity from CTTSs (see Table~\ref{twhyaaumic}), 
based on direct detections from a few objects \citep{Herczeg2004,Yang2011,France2011} and 
indirect detections from strong H$_2$ lines \citep[e.g.][]{Herczeg2006,Ingleby2009}.  The total stellar Ly$\alpha$ flux is not directly observable 
because neutral hydrogen in the ISM and circumstellar environment scatters Ly$\alpha$ photons out of our line of sight to the star.  
The Ly$\alpha$ line is also not covered by the ACS/SBC PR130L prism. 
For WTTSs, correlations with $L_X$ provided by \citet{Wood2005} could be used to infer Ly$\alpha$ luminosities.
For both CTTSs and WTTSs, reconstruction of the stellar Ly$\alpha$ line is needed to properly estimate the Ly$\alpha$ flux seen by the circumstellar
gas.

\subsection{Evolution of FUV Emission for PMS Stars}
\begin{figure}[ht]
  \begin{center}
    \includegraphics[scale=0.65]{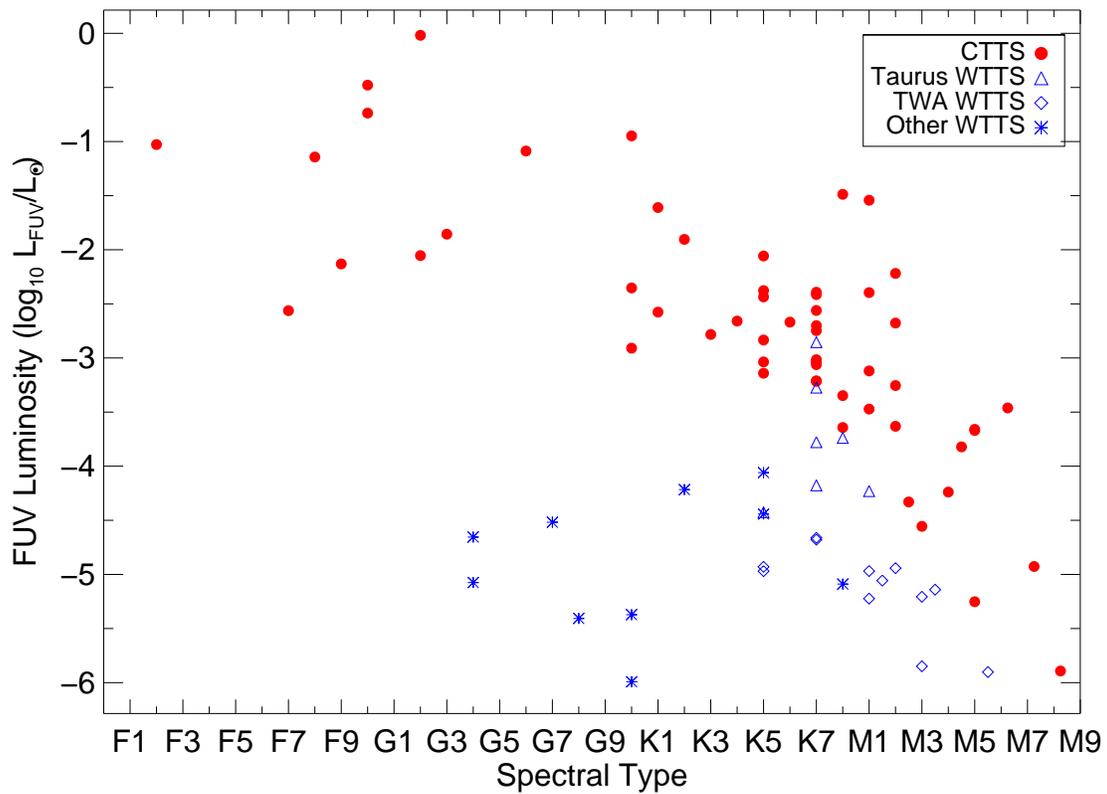}
       \caption{ FUV luminosity is plotted against spectral type.
           CTTSs are plotted as red filled circles. WTTSs in different regions are plotted in blue as different symbols.
           The triangles are WTTSs in Taurus, the diamonds are WTTSs in TWA, and the asterisks represent WTTSs in other regions. }
           \label{fuvlumsptype}
              \end{center}
    \end{figure}

\begin{figure}[ht]
  \begin{center}
    \includegraphics[scale=0.65]{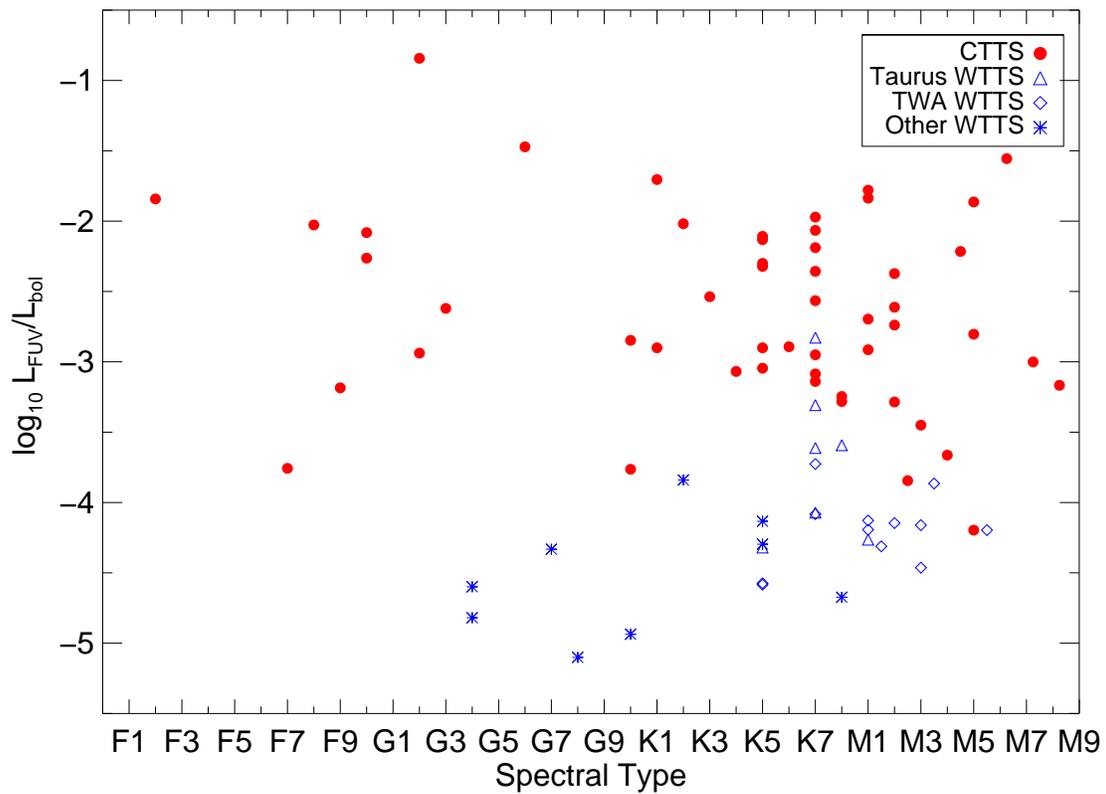}
       \caption{ Ratios of FUV luminosity to bolometric luminosity are plotted against spectral types.
           CTTSs are plotted as red filled circles. WTTSs in different regions are plotted in blue as different symbols.
           The triangles are WTTSs in Taurus, the diamonds are WTTSs in TWA, and the asterisks represent WTTSs in other regions. }
           \label{lfuvlbolsptype}
              \end{center}
    \end{figure}

Figures~\ref{fuvlumsptype} and~\ref{lfuvlbolsptype} show the FUV luminosity, $L_{\rm{FUV}}$, and its ratio to the bolometric luminosity, 
$L_{\rm{FUV}}/L_{{bol}}$, versus spectral type, respectively. 
For stars with spectral type later than K0, where the extinction estimates are more accurate, 
the FUV luminosity, $L_{\rm{FUV}}$, of CTTSs is $\sim 10^{-2.75} L_{bol}$, with a scatter of $\sim 0.65$ dex, and 
the FUV luminosity of WTTSs is $\sim 10^{-4.06} L_{bol}$ with a scatter of $\sim 0.43$ dex. The higher FUV luminosities of CTTSs are due to accretion 
and related processes. Many of the WTTSs in our sample are thought to be older than their CTTS counterparts and have smaller bolometric luminosities and, 
consequently, even lower absolute FUV luminosities. The WTTSs in Taurus generally show stronger FUV emission than the WTTSs of similar
spectral types in TWA and other regions. Our sample is biased in age and mass. Many of the K stars are from the $\sim 2$ Myr old \citep{Palla2000}
Taurus Molecular Cloud, and all M dwarfs are located in the 10 Myr \citep{Barrado2006} TWA, while G stars are typically older field dwarfs.

\citet{Ingleby2011} showed for their sample that the FUV luminosity between 1230 and 1800 \AA\ correlates well with $L_{acc}$ and both quantities decrease
with age from 1 Myr to 1 Gyr.
As shown in Figure~\ref{accretionlfuv}, despite the scatter, we also find that the FUV emission of CTTSs clearly scales with the accretion rate.
The highest accretion rates likely occur in outbursts during the Class 0 and I stages \citep{Dunham2010}, when the accretion and 
any FUV emission are mostly or entirely obscured from our view by dense envelopes. During these stages, the mean accretion rate 
is likely higher than measured for CTTSs, with a correspondingly higher FUV luminosity. The CTTS stage includes stars with accretion 
rates of that range over 2 orders of magnitude at a given mass, with some suggestion that older CTTSs have lower accretion rates. 
As the accretion rate decreases, so does the FUV luminosity.  The chromospheric emission provides the lower limit on the strength 
of FUV emission from young stars. 


\subsection{Importance of Far-UV Spectra for Photoevaporation and 
Photochemistry of Close-in Exoplanet Atmospheres}

The discovery and characterization of Jupiter mass exoplanets close to their 
host stars by ground-based telescopes and the CoRoT and Kepler missions 
raises many questions concerning the photochemistry of their atmospheres and
mass loss due to hydrodynamic and magnetohydrodynamic processes. For example,
the well studied exoplanet HD~209458b, which has a semimajor axis of 0.045 AU
and a 3.52 day orbital period about its G0~V host star \citep{Knutson2007}, 
receives a radiative flux 13,400 times that of Jupiter. Like many other hot Jupiters, HD~209458b
is inflated with a radius 1.25 times that of Jupiter, which could be due in 
part to external heating. While the visual and near-UV (NUV) stellar flux  
dominates the external heating of the lower atmospheres of hot Jupiters, the 
stellar FUV, EUV, and X-ray fluxes will control ionization and dissociation processes in the 
outer atmosphere and drive mass loss through heating of the 
exosphere.

Models of hydromagnetic mass loss from hot Jupiters \citep[]{Yelle2004, Tian2005,GarciaMunoz2007, Murray-Clay2009} 
assume that the mass loss is driven by absorption by H$_2$ and \ion{H}{1} in the upper atmopshere.
The inclusion of photoabsorption of UV photons by C, N, and O can greatly increase the 
mass loss rate \citep{GarciaMunoz2007}.   The FUV emission can also affect 
atmospheric chemistry by photodissociation of many molecules, including H$_2$, H$_2$O, and  CH$_4$.  Biologically-important 
O$_2$ is dissociated by $\lambda < 2500$~\AA\ radiation leading to the 
formation of O$_3$. 


Realistic assessments of the FUV flux must be included in theoretical models to explain the large mass flux for HD~209458b \citep{Linsky2010} 
and to accurately describe the photochemistry in the atmospheres of young or forming planets. During the active accretion phase, 
a Jovian-mass planet will migrate inward to where the disk is truncated.  In this phase, the optically thick disk may shield the planet 
from direct irradiation from the central star. Once the disk dissipates, the planets see bright emission from the central star.
\citet{Ribas2005} have collected the available UV and X-ray fluxes (between 1 and 1200 \AA) as a function of age for solar-mass stars older than 50 Myr and 
provided scaling laws for the age dependence of the flux in different bands. They found that the emission in all wavelength intervals decreases by 
a few orders of magnitude following power laws from 0.1 to 7 Gyr, and higher energy emissions display steeper decreases.
\citet{Ingleby2011} extended this study to younger stars and found that the \citet{Ribas2005} power laws are valid for stars between 15 and 100 Myr
for X-ray and FUV (1230--1800 \AA\ bandpass) luminosities. The FUV emission of WTTSs in our sample have similar strengths as 
that reported in \citet{Ingleby2011} and thus a few orders of magnitude stronger than the much older solar analogs in \citet{Ribas2005}.

The above analysis excludes Ly$\alpha$ emission, which dominates the total FUV emission from both WTTSs and CTTSs \citep{Wood2004,Herczeg2004}.  For CTTSs, this emission can be scattered deeper into the disk than other FUV photons \citep[Herczeg 2006]{Bethell2009}.  However, most planets are located beyond the dust sublimation radius and will be shielded from any optically-thick radiation from the disk.  Once the disk dissipates, any hot Jupiter will see the total FUV emission from the star, including Ly$\alpha$.  The incident radiation field for WTTSs should include both the 1250--1700 \AA\ emission discussed here and the Ly$\alpha$ radiation, which can be estimated from the stellar X-ray luminosity \citep{Wood2005}.

\subsection{HD 98800 N: a debris disk or an accretion disk?}

\begin{figure}[ht]
  \begin{center}
    \includegraphics[scale=0.65]{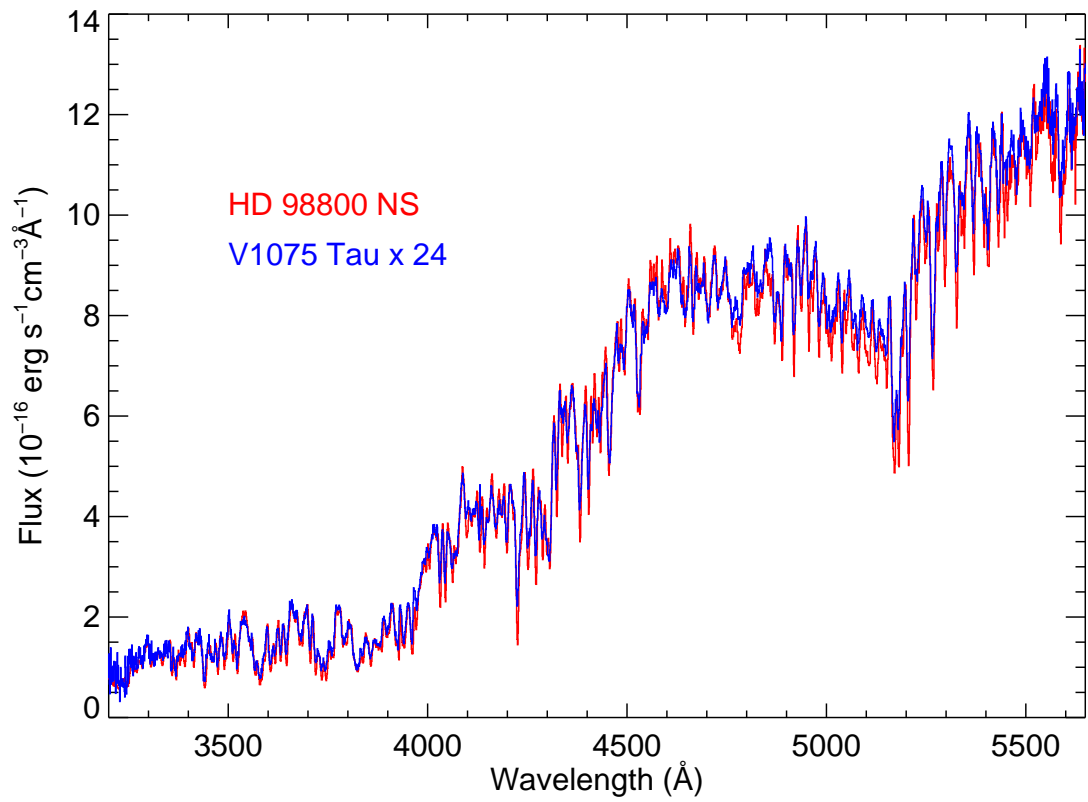}
       \caption{The optical spectra of HD 98800 and V1075 Tau.}
           \label{hd98800}
              \end{center}
    \end{figure}

HD 98800 (K5) is a quadruple system that consists of two spectroscopic binaries separated by $0\farcs8$ \citep{Boden2005}.  
An IR excess from the N component indicates the presence of a dust disk \citep{Gehrz1999,Koerner2000,Prato2001}.  
Both the IR spectral energy distribution and high-resolution sub-mm images indicate a hole in the dust distribution in the disk, 
with an inner truncation radius of 3.5--6 AU from the central binary \citep{Furlan2007,Akeson2007,Andrews2009}. 
A low-resolution 3300--5500 \AA\ spectrum (Figure~\ref{hd98800}) that includes both components shows no evidence for accretion, 
with $\log \dot{M}<-9.9$ (or $\log \dot{L}/L_\odot<-2.5$), following \citet{Herczeg2008}.  
\citet{Salyk2009} did not detect any CO rovibrational emission in a high-resolution M-band ($\sim$ 5 $\mu$m) spectrum of HD 98800 N (ID: 63).  
Prior to this work, the absence of any gas or accretion signatures suggested that the disk should have been classified as a debris disk rather than a transition (or cold) disk with ongoing accretion.

\begin{figure}[ht]
  \begin{center}
    \includegraphics[scale=0.65]{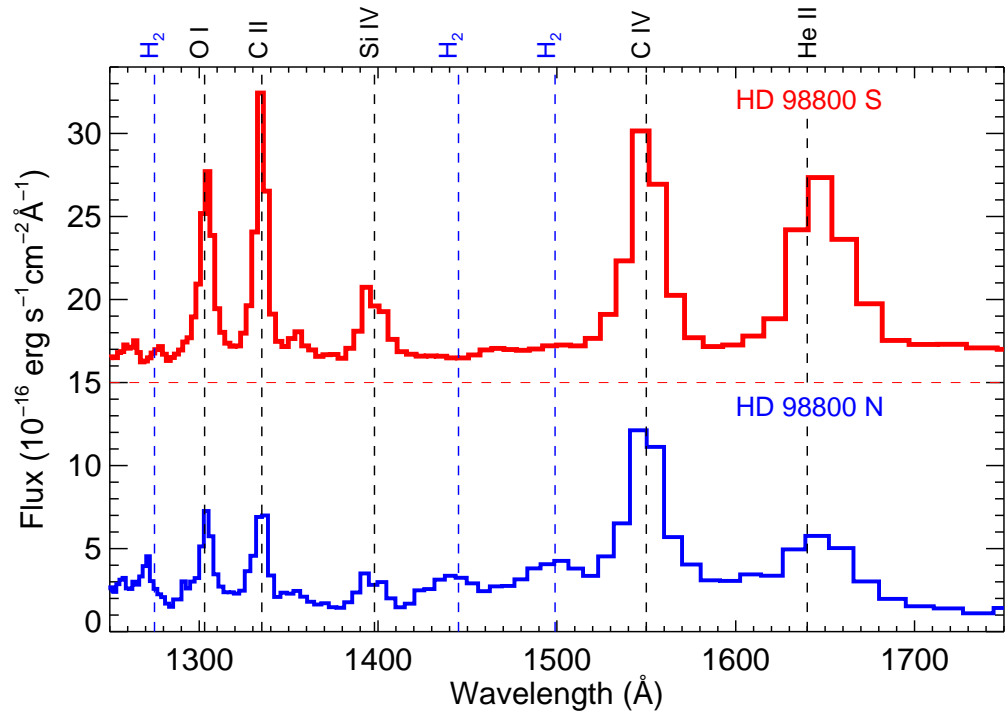}
       \caption{Spectra of HD 98800 N an S are shown. The spectrum of HD 98800 S is shifted by $+1.5 \times 10^{-15}$ erg s$^{-1}$ cm$^{-2}$ \AA$^{-1}$ 
          for better viewing. Strong atomic emission lines are labeled, and a few H$_{2}$ emission features shown in HD 98800 N are also identified.}
           \label{hd98800_h2}
              \end{center}
    \end{figure}

The emission at 1420 and 1500 \AA\ from HD 98800 N is Ly$\alpha$-pumped H$_2$ emission (Figure \ref{hd98800_h2}). 
The general rise shortward of 1700 \AA\ is likely explained, at least in part, by a combination of Ly$\alpha$-pumped H$_2$ emission 
and H$_2$ emission following a collision with energetic electrons \citep{Herczeg2002,Bergin2004,France2010b}.  
\citet{Ingleby2009} detected this H$_2$ emission in the FUV spectrum of every classical T Tauri star but not for any star with a debris disk.  
FUV H$_2$ line emission has been previously detected around one debris disk, AU Mic \citep{France2007}.  
However, that emission produced in cold H$_2$ gas is faint, and could not explain the H$_2$ emission seen in the PR130L spectrum.

Our detection of bright FUV H$_2$ emission from HD 98800 N demonstrates that warm gas persists in its disk. This detection suggests that 
HD98800 N has ongoing accretion with a rate below previous detection limits.  Ingleby et al.~(2011) describes a similar case, RECX 11, 
which lacks any detectable excess U-band emission but has accretion detectable in several diagnostics, including the FUV.

\subsection{Evidence for Depletion of Refractory Elements in Accretion Flows}\label{refractory}
\begin{figure}[ht]
  \begin{center}
    \includegraphics[scale=0.65]{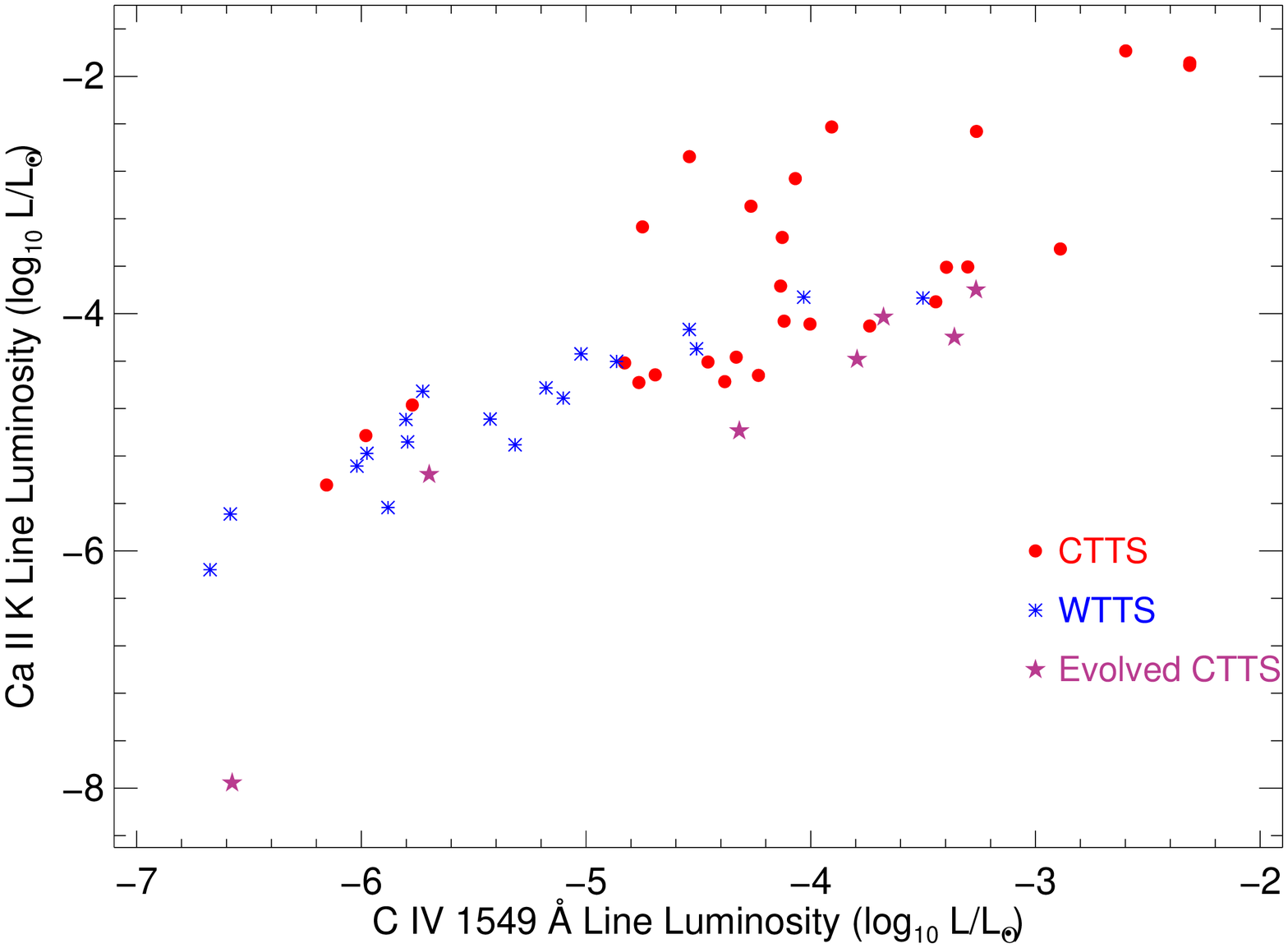}
       \caption{ The \ion{Ca}{2} K line luminosity is plotted against the \ion{C}{4} $\lambda$1549 doublet luminosity for 
              WTTSs (blue asterisks), CTTSs (red filled circles), and evolved CTTSs (purple stars) as defined in \S \ref{refractory}.  }
           \label{evolved}
              \end{center}
    \end{figure}

Figure~\ref{evolved} compares the \ion{Ca}{2} K and \ion{C}{4} line luminosities for WTTSs, CTTSs, and more evolved CTTSs. 
As defined here, these more evolved CTTSs include members of the TW Hya and Upper Sco Associations, the cold (or transitional) disks GM Aur, DM Tau, 
and LkCa 15, the settled disk V836 Tau, and SCH J0518, which is located off of the main Taurus fields in a region with a low disk fraction.

For WTTSs, the \ion{Ca}{2} and \ion{C}{4} lines are both produced by chromospheric emission, while for CTTSs, these lines are produced 
by a combination of chromospheric emission and accretion-related processes. The correlation between \ion{Ca}{2} and \ion{C}{4} line luminosities 
observed at different times is very tight for WTTSs, indicating that variability is moderate. Generally, extinctions for WTTSs are lower and less
uncertain than for CTTSs, and that may contribute to the tighter relationship between \ion{Ca}{2} and \ion{C}{4} line emission.
Larger scatter is expected and observed for CTTSs because accretion can be highly variable and the extinctions are more uncertain.

The more evolved CTTSs almost all have low \ion{Ca}{2}/\ion{C}{4} luminosity ratios relative to most other CTTSs in our sample. The exception is TWA 3N (ID: 59), which is only weakly 
accreting \citep{Herczeg2009} at a low enough rate that emission in both lines could be dominated by the chromosphere.  
The low \ion{Ca}{2}/\ion{C}{4} line ratios could be explained by the preferential depletion of refractive metals into grains that have settled 
into a dead zone below the surface layers in the disk and do not take part in accretion \citep{Gammie1996}.  
For the 10 Myr old CTTSs, TW Hya and 2M1207 (TWA 27), 
the weakness in FUV Si lines and, for 2M1207, \ion{Mg}{2} h \& k lines, relative to C lines has been interpreted as possible evidence 
for Si depletion into dust grains in the disk \citep{Herczeg2002,France2010}. 
This interpretation has some support from high-resolution X-ray spectroscopy of more evolved CTTSs 
that show anomalously high Ne/Fe and Ne/O abundance ratios, with line ratios suggesting an origin 
in the accretion flow \citep{Kastner2002,Stelzer2004,Guenther2007}.  Similarly, certain metals can be depleted in outflows \citep[e.g.][]{Nisini2005,GarciaLopez2010}, which may indicate that the gas in the wind launch region is similarly depleted in refractory elements.

The low \ion{Ca}{2}/\ion{C}{4} ratios for the cold disks in Taurus support the idea that refractive metals 
are also depleted in the accretion flow in these disks. 
The disks have undergone some evolution relative to the typical disks seen in Taurus. 
The depletion of Ca may suggest that refractive metals have settled into large grains or planetessimals.  
This analysis is restricted to Ca because the Si lines are difficult to separate from H$_2$ emission in the low-resolution ACS PR130L spectra.

\subsection{Spatially Extended Jets in the FUV}

H$_2$ emission from young stars can be spatially extended and related to outflows, as is seen prominently around T Tau 
\citep[see discussion in \S\ 3.1; ][]{Walter2003,Saucedo2003}.  
The stellar emission strengths of the \ion{Si}{4} and \ion{C}{4} lines are similar in both observations,
but the larger {\it IUE} aperture detects much stronger H$_2$ emission, as expected for spatially extended emission (see Figure~\ref{aperture}).

In most spectra presented here, the H$_2$ emission is centered on-source, with no significant contribution from spatially extended outflows.  
The lack of spatially extended emission does not necessarily indicate that the H$_2$ emission traces the disk. 
High angular-resolution spectra of RU Lup, T Tau, DG Tau, and DF Tau show that the on-source emission includes 
at least some emission from an outflow \citep{Herczeg2006}.  Off-source H$_2$ emission is detected from two other sources, DP Tau and 2MASS J04141188+2811535.
An asymmetric bipolar jet (HH 231) is associated with DP Tau \citep{Mundt1998}, so results from the DP Tau spectrum are not included 
in any of the correlations presented in \S 3.2.  
Near-IR AO images of 2MASS J04141188+2811535 indicate that the object lacks a near-IR companion of similar brightness (Adam Kraus, private communication).  The secondary component is likely an H$_2$ jet, which is also consistent with the presence of a  weak emission between the primary and secondary positions and with the presence of bright optical forbidden lines \citep{Herczeg2008}.


\section{SUMMARY}

In this paper, we present a FUV spectral atlas of 91 CTTSs and WTTSs, as observed by the GHRS, STIS, and ACS instruments on {\it HST}, including
some observations by \emph{IUE} and \emph{FUSE}.  
We find that FUV line luminosities are well correlated with each other over a large range of luminosity.  
The same correlation applies to both CTTSs and WTTSs, which indicates that the emission measure has a similar distribution versus temperature 
for chromospheric activity and for accretion.  Accretion significantly increases the strength of these lines and the total FUV luminosity, 
with $\log \frac{L_{\rm{FUV}}}{L_{bol}} \sim -2.7$
for CTTSs and $\sim -4.0$ for WTTSs.
For stars with evolved but accreting disks, including transition disks, the \ion{Ca}{2} H \& K lines are weaker than expected, 
which supports the suggestion that refractory elements may be depleted in accretion flows onto these stars.
We provide easy-to-use prescriptions to obtain FUV luminosities from either optical line luminosities or from a known accretion rate.  
We also discuss several individual spectra, including the detection of warm H$_2$ emission from HD 98800 N, 
which suggests ongoing accretion for a disk that had been previously classified as a debris disk.



\appendix

\section{Tight Binaries in the UV}\label{binary}

\subsection{EZ Ori}
The STIS acquisition image of EZ Ori reveals two components.  
The image was obtained with the STIS CCD $(0\farcs0507$ pixels) and the F28x500 [\ion{O}{3}] filter, which is centered at 5007 \AA\ and covers $\sim 12$ \AA. 
The secondary is separated by $\sim0\farcs18$ from the primary with a PA$\sim76^\circ$ and a flux in the bandpass equal to $\sim 8$\% of the primary.  
The FUV and NUV spectra were obtained with a large slit that includes both stars. The extracted FUV and NUV spectra are consistent with the presence 
of a binary at that position, with the wavelength solution of the secondary shifted by four pixels because it was not centered in the 
wide ($2^{\prime\prime}$) slit. About 20\% of the flux in FUV emission lines is from EZ Ori B, although \ion{C}{2} emission is relatively weak. 
Also, 15\% of NUV emission and 10\% of the \ion{Mg}{2} 2800 \AA\ doublet is from EZ Ori B. Both components are likely accreting based on the strength 
of these lines and the NUV continuum. For purposes of this paper the two sources are treated as a single star.

\subsection{TWA 16}
The acquisition image of TWA 16 was obtained with the F150LP long-pass filter.
TWA 16 is resolved into two stars separated by $0\farcs56$ at a PA=316$^\circ$ and with a brightness difference of 0.095 mag.

\subsection{UZ Tau E}
UZ Tau was acquired with the F165LP long-pass filter, which covers spectral regions dominated by the continuum emission.  
In this image, the northern component is stronger. 
The two components of UZ Tau E have a flux ratio of $\sim 1.95$ and are separated by 0\farcs32 mas at a PA=7.9. 
The two components of UZ Tau E are aligned close to the dispersion direction and are not resolved in the prism spectrum.  H$\alpha$ measurements from \citet{Hartigan2003} indicate that both components of UZ Tau E are accreting. 

\subsection{2M1614}
\citet{Metchev2009} found 2MASS J16141107-2305362 (2M1614, or PZ 161411-230536) to be a nearly equal-mass binary 
separated by $0\farcs222$ with a PA=304.8$^\circ$. We confirm the presence of two components, but find a separation of $0\farcs309$ with a PA=296.7$^\circ$.  
The difference could be due to orbital motion.
Accretion has been detected in unresolved optical spectra that includes both objects \citep{Pascucci2007}.  The presence of strong H$_2$ emission from the N component and the lack of detectable H$_2$ emission from the S component suggests that the system is a CTTS+WTTS binary.

\section{Uncertainties In Data From Different Instruments}\label{instrument}   
\begin{figure}[ht]
  \begin{center}
    \includegraphics[scale=0.70]{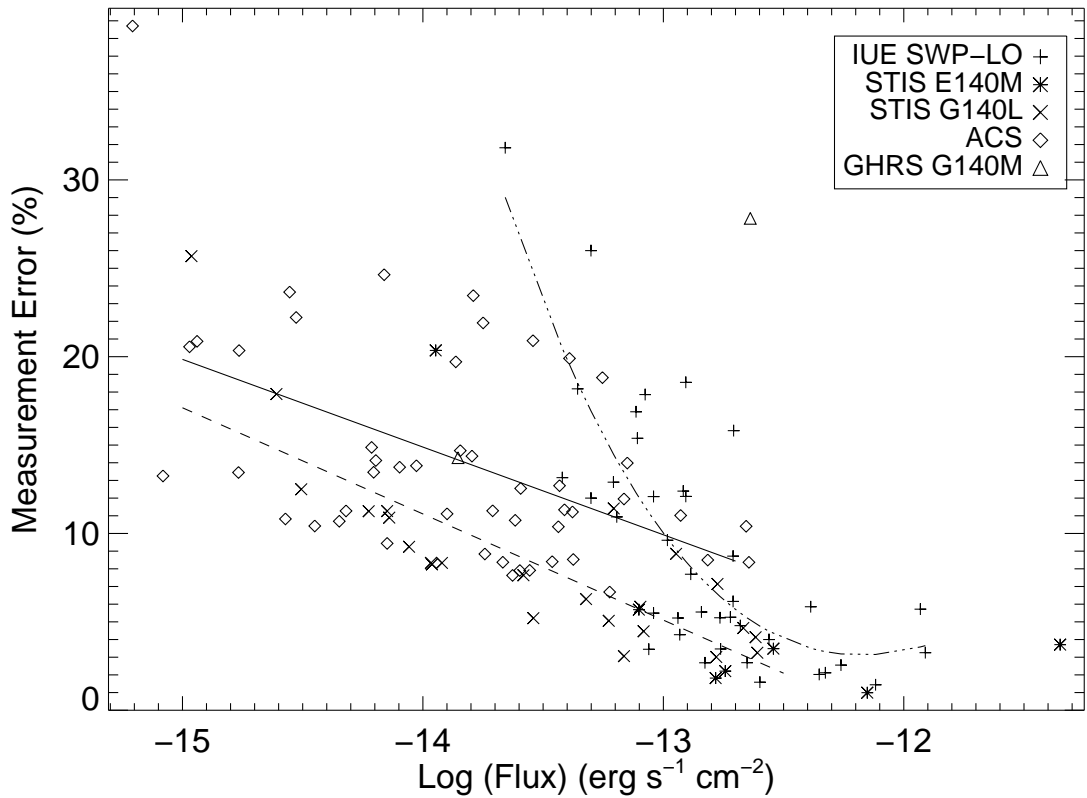}
       \caption{ The percentage errors in flux measurements are plotted against fluxes for the \ion{C}{4} doublet listed
         in Table 4 of this work and the IUE data listed in Table 6 of \citet{Valenti2000}. The different data sets are indicated by 
         different symbols noted in the legend. The solid line is a 
         linear least-squares fit to the ACS data, and the dashed line is a fit to the
         STIS G140L data. The curved line is a rough fit to the IUE data.}
           \label{c4compare}
              \end{center}
    \end{figure}

We have presented PMS star spectra obtained from four instruments on
HST with different spectral resolutions, sensitivities, and measurements 
uncertainties. 
Figure~\ref{c4compare} compares the \ion{C}{4} 1549~\AA\ doublet fluxes and percent errors 
of PMS stars measured by the four HST instruments and the IUE SWP-LO mode. The errors include both statistical errors
in the spectra and flux measurement errors.
The faintest PMS stars were 
observed by the instruments with the lowest spectral resolution (ACS and 
STIS G140L). For ACS the percent errors increased from typically 10\% 
at high flux levels to $>20$\% at its flux limit of $10^{-15}$ ergs cm$^{-2}$ s$^{-1}$. 
Note, however, the large scatter in the errors relative to the least squares
linear trend line. For STIS G140L observations the errors are roughly 5\% 
at high flux levels increasing to 
$>20$\% at the lowest flux levels, and the scatter about the linear trend 
line is relatively small. The moderate-resolution STIS E140M grating
has provided fluxes with 5\% or smaller errors for high flux level sources 
increasing to 20\% for the lowest flux source at $10^{-14}$ ergs cm$^{-2}$ s$^{-1}$.
As the curved trend line shows, \ion{C}{4} spectra obtained with the IUE's SWP-LO 
camera typically show 5\% flux errors for sources with high flux levels
($>10^{-13}$ ergs cm$^{-2}$ s$^{-1}$) but provide no useful data at flux levels
a factor of 3 lower. Table~\ref{instruments} includes a rough estimate of the \ion{C}{4}
flux levels for each instrument, below which measurement errors are generally
greater than 10\%.


\bibliographystyle{apj}
\bibliography{atlasref}

\pagestyle{empty}



\begin{deluxetable}{lrllcccccccr}
\rotate
  \tablewidth{0pt}
  \tablecaption{ Summary of Observations. \label{obslog}}
\tabletypesize{\scriptsize}
\tablehead{
\colhead{ID\tablenotemark{a}}& \colhead{HBC\tablenotemark{b}}
&\colhead{Name}&\colhead{Alternate Name} 
&\colhead{$\alpha$(2000)} & \colhead{$\delta$(2000)} & \colhead{Instrument}  &\colhead{UT Date}
&\colhead{N\tablenotemark{c}}&\colhead{$\emph{t}_{\rm exp}$ (s)\tablenotemark{d}}& \colhead{Program ID\tablenotemark{e}} 
}
\startdata
1 &...&DK Cet   &HD 12039    &01 57 48.98 &$-21$ 54 05.3& ACS PR130L& 2007 JUL 13 &1&  2440.0   &  10810\\ 
2 & 20&LkH$\alpha$ 330 &...  &03 45 48.28 & +32 24 11.9 & ACS PR130L& 2007 SEP 25 &1&  1786.0   &  11151\\
3&...&2M0414\tablenotemark{f}&XEST 20-OM-002&04 14 11.98&+28 12 00.2&ACS PR130L&2008 NOV 08&1&1769.0 &11151\\ 
4 & 23&FM Tau   &Haro 6-1    &04 14 13.44 & +28 12 51.1 & ACS PR130L& 2007 OCT 31 &1&  2482.0   &  10840\\ 
5 & 26&FP Tau   &\mha\ 259-23&04 14 47.00 & +26 46 26.5 & ACS PR130L& 2007 AUG 15 &1&  2482.0   &  10840\\ 
6 & 27&CX Tau   &\mha\ 259-22&04 14 47.86 & +26 48 11.0 & ACS PR130L& 2007 AUG 08 &1&  2482.0   &  10840\\
7 &368& LkCa 3  &V1098 Tau   &04 14 47.94 & +27 52 35.8 & STIS G140L& 2003 SEP 20 &1&  1859.2   &   9790\\ 
8 &370& LkCa 4  &V1068 Tau   &04 16 28.04 & +28 07 35.9 & STIS G140L& 2004 MAR 01 &1&  1859.2   &   9790\\
\enddata
\tablenotetext{a}{Internal identification number for cross-referencing.}
\tablenotetext{b}{Catalog number from \citet{herbig1988}.}
\tablenotetext{c}{Number of exposures.}
\tablenotetext{d}{Exposure time.}
\tablenotetext{e}{HST and IUE program ID's.}
\tablenotetext{f}{Name Abbreviations: 2M0414 for 2MASS J04141188+2811535; 2MJ0439 for 2MASS J04390163+2336029; 2MASS J04442713+2512164;
  2M0518 for 2MASS J05180285+2327127; 2M1111 for 2MASS J11114632-7620092; 2M1207A for 2MASS J12073347-3932540; 2M1614N for 2MASS J16141107-2305362 N; 
  2M1614S for 2MASS J16141107-2305362 S; 2M1842 for 2MASS J18425797-3532427; 2M1852 for 2MASS J18521730-3700119.}
\tablenotetext{g}{Multiple program IDs: MR381, FSCJL and UK353.}
\end{deluxetable}

\begin{deluxetable}{llccccrrrrrrc}
\rotate
 \tablewidth{0pt}
 \tablecaption{ Pre-Main Sequence Stars. \label{pmstable}}
\tabletypesize{\scriptsize}
\tablehead{
\colhead{ID} &\colhead{Name}&\colhead{Region} 
&\colhead{Distance (pc)} & \colhead{Type\tablenotemark{a}} & \colhead{Category}  &\colhead{$A_{V}$}
&\colhead{log$L_{phot}$\tablenotemark{b}}& \colhead{log$L_{acc}$\tablenotemark{c}}  
& \colhead{log$L_{\rm{X}}$\tablenotemark{d}}&\colhead{H$\beta$\tablenotemark{e}} &\colhead{Ca II K\tablenotemark{e}}
&\colhead{Reference \tablenotemark{f}}
}
\startdata                                
1 &DK Cet          &   field      & 42& G4    &WTTS& 0.0& -0.05&...   &...   &  ...  & ...  & 1,2 \\ 
2 &\Lkha 330       &   Per        &250& G3    &CTTS& 1.8&  1.2 &...   &...   &  ...  & ...  &  3 \\
3 &2M0414          &   Taurus     &140& M6.25 &CTTS& 0.7& -1.90& -2.50&...   &  2.50 & 1.90 & 4 \\ 
4 &FM Tau          &   Taurus     &140& M1    &CTTS& 0.3& -0.77& -1.7 &29.69 &  17.8 & 3.32 &5,33 \\
5 &FP Tau          &   Taurus     &140& M4    &CTTS& 0.0& -0.57& -2.8 &...   &  10.7 & 1.54 &  5 \\
6 &CX Tau          &   Taurus     &140& M2.5  &CTTS& 0.2& -0.48& -2.6 &...   &  14.9 & 2.12 &5 \\
7 &LkCa 3          &   Taurus     &140& M1    &WTTS& 0.4&  0.04& ...  &...   &  2.42 & 4.35 &         6 \\ 
8 &LkCa 4          &   Taurus     &140& K7    &WTTS& 1.2& -0.02&...   &30.30 &  5.86 & 4.30 & 5,34 \\
\enddata
\tablenotetext{a}{``SB'' stands for spectroscopic binary and ``B'' stands for binary.}
\tablenotetext{b}{Photospheric luminosity in units of solar luminosity.}
\tablenotetext{c}{Accretion luminosity in units of solar luminosity.}
\tablenotetext{d}{X-ray (0.3 - 10.0 keV) luminosity in units of erg~$\rm{s^{-1}}$.}
\tablenotetext{e}{Fluxes of the H$\beta$ and \ion{Ca}{2} K lines in units of $10^{-14}$ erg~$\rm{s^{-1}cm^{-2}}$.}
\tablenotetext{f}{References:  1:  \citet{White2007};  2:  \citet{Perryman1997}; 3:  \citet{CohenKuhi1979};  }
\tablenotetext{~}{4:  \citet{Herczeg2008};  5:  \citet{White2001}; 6:  \citet{Kenyon1995};  }
\tablenotetext{~}{7:  \citet{Gullbring1998}; 8:  \citet{Nguyen2009};  9:  \citet{Valenti1993};}
\tablenotetext{~}{10:  \citet{Gullbring2000}; 11:  \citet{Calvet2004};  12:  \citet{GarciaLopez2010};}
\tablenotetext{~}{13:  \citet{Santos2005};  14:  \citet{Mamajek2005}; 15:  \citet{Webb1999};}
\tablenotetext{~}{16:  \citet{Valenti2005};  17:  \citet{Herczeg2009}; 18: \citet{Sterzik1999};}
\tablenotetext{~}{19:  \citet{Ducourant2007}; 20:  \citet{Zuckerman2001};  21:  \citet{Mamajek2002};}
\tablenotetext{~}{22:  \citet{Herczeg2005}; 23: \citet{dezeeuw1999};  24:  \citet{Walter1994};}
\tablenotetext{~}{25:  \citet{Preibisch1999};  26:  \citet{Pascucci2007}; 27:  \citet{Bouvier1990};}
\tablenotetext{~}{28:  \citet{Hughes1994};  29:  \citet{Loinard2008}; 30:  \citet{Donati2007};}
\tablenotetext{~}{31: \citet{Basri1990}; 32:  \citet{Neuhauser2008}; 33: \citet{Guedel2007};} 
\tablenotetext{~}{34: \citet{BrownHEAD2010}; 35: this paper from Swift data; 36: this paper from Chandra data. }
\end{deluxetable}

\begin{deluxetable}{lrlcccccccrc}
\rotate
  \tablewidth{0pt}
  \tablecaption{ Summary of FUV Instruments. \label{instruments}}
\tablehead{
\colhead{Observatory} & \colhead{Instrument}
&\colhead{Setting}& \colhead{Wavelength Range} &\colhead{Aperture}
&\colhead{Resolution} &\colhead{log $F_{\rm{C~IV}}$\tablenotemark{a}}
}
\startdata      
HST & STIS    & E140M  & 1150--1700 & $0.2\arcsec \times 0.06\arcsec$ & 45,800   &  -13.2\\
HST & STIS    & G140L  & 1150--1700 &  $52\arcsec \times 2\arcsec$    & 935--1440&  -14.0\\
HST & ACS/SBC & PR130L & 1250--2000 &  $28\arcsec \times 31\arcsec$   & 60--150  &  -12.7\\
HST & GHRS    & G160M  & 1382--1420 &  $2\arcsec \times 2\arcsec$     & 18,000   &   N/A \\
    &         &        & 1532--1568 &  $2\arcsec \times 2\arcsec$     & 19,000   &   N/A \\
IUE &         & SWP-LO & 1150-1975  & $10\arcsec \times 20\arcsec$    &$\sim 300$&  -13.0\\
\enddata
\tablenotetext{a}{C IV flux levels below which the measurement errors are typically $\ge$ 10\%.}
\end{deluxetable}

\begin{deluxetable}{llllcccccccc}
\rotate
  \tablewidth{0pt}
  \tablecaption{ Line Flux Measurements. \label{allflux}}
\tabletypesize{\scriptsize}
\tablehead{
\colhead{ID} &\colhead{Name} & \colhead{Instrument} &\colhead{\rm{O I 1304}}&\colhead{C II 1335}
   &\colhead{O I 1355}&\colhead{Si IV 1394}&\colhead{Si IV 1403}&\colhead{C IV 1549}
   &\colhead{He II 1640} &\colhead{ $\rm{H_{2}}$ emission } &\colhead{$L_{\rm{FUV}}$}
}
\startdata                         
1 &DK Cet              &ACS       &10.3(1.5) & 13.4(1.7)&...       & 16.9(2.9)& ---      &40.7(8.1) &28.0(8.5) & No &$2.21 \times 10^{-5}$\\
2 &\Lkha 330           &ACS       &   ...    &   ...    &...       &   ...    & ...      &0.83(0.11)&   ...    & No &$3.61 \times 10^{-4}$\\
3 &2M0414              &ACS       &0.94(0.20)&2.50(0.38)&...       &4.16(0.76)& ---      &17.8(3.9) &4.16(0.76)& No &$3.46 \times 10^{-4}$\\
4 &FM Tau              &ACS       &...       &...       &...       &...       &...       &16.2(3.8) &21.4(7.0) & No &$3.37 \times 10^{-4}$\\
5 &FP Tau              &ACS       &0.48(0.08)&1.22(0.13)&...       &1.91(0.20)& ---      &1.72(0.35)&...       & Yes&$5.76 \times 10^{-5}$\\
6 &CX Tau              &ACS       &0.41(0.08)&0.83(0.13)&...       & ...      &...       &1.71(0.23)&...       & Yes&$4.67 \times 10^{-5}$\\
7 &LkCa 3              &STIS G140L&3.85(0.42)&3.41(0.32)&0.46(0.16)&1.03(0.19)&0.63(0.17)&5.95(0.67)&3.89(0.75)& No &$5.89 \times 10^{-5}$\\
8 &LkCa 4              &STIS G140L&7.31(0.54)&10.3(0.6) &1.10(0.26)&4.39(0.41)&2.79(0.36)&28.8(1.5) &12.6(1.4) & No &$1.40 \times 10^{-3}$\\
\enddata
\tablecomments{The flux values are in units of $10^{-15}$ erg~$\rm{s^{-1}cm^{-2}}$. ``...'' indicates that the line
    is not apparent in the spectrum. ``---'' indicates that the Si IV 1403 line is not resolved and 
    the flux in the Si IV 1394 column represents the total flux of the doublet. The total FUV luminosity between 1250 and 1700\AA, $L_{FUV}$,
    are in solar units.}
\end{deluxetable}

\begin{deluxetable}{lccrcrcrcr}
  \rotate
  \tablewidth{0pt}
  \tablecaption{FUV line fluxes extracted from synthetic ACS PR130L spectra and their corresponding pertcent differences to 
           the line fluxes in the actual STIS G140L Data.\label{pseudoacs}} 
  \tabletypesize{\footnotesize}
\tablehead{
\colhead{ID}  &  \colhead{Object}&\colhead{O I 1304}& \colhead{\%} & \colhead{C II 1335}& 
\colhead{\%}  &\colhead{Si IV 1394 + Si IV 1403}& \colhead{\%}  &\colhead{C IV 1549}&\colhead{\%}
}
\startdata
7   &LkCa 3     & 3.91   &  1.6  &  3.56         & 4.3 &  1.69                           & 1.8 &    6.90        & 16.0 \\
8   &LkCa 4     & 7.21   & -1.4  &  10.0         &-2.9 &  7.46                           & 3.9 &    32.1        & 11.5 \\
10  &BP Tau     & 109    &  5.3  &  45.0         &26.7 &  64.7                           & 9.6 &    247         &  0.5 \\
11  &LkCa 7     & 4.51   &  2.5  &  5.21         &-6.1 &  2.28                           & 3.6 &    12.5        &  4.2 \\
13  &RY Tau     & 71.7   & -5.0  &  45.4         &-0.8 &  76.8                           &-10.3&    698         &  3.9 \\
14  &T Tau      & 253    & -1.6  &  31.2         &-17.7&  108                            &  0.0&    225         &  4.7 \\
18  &V827 Tau   & 4.43   &  4.2  &  4.55         & 4.1 &  2.25                           &-17.6&    12.7        & 17.2 \\
19  &V826 Tau   & 2.62   & -0.4  &  4.24         & 0.7 &  2.16                           &-8.5 &    11.3        &  3.7 \\
\enddata
\tablecomments{The flux values are in units of $10^{-15}$ erg~$\rm{\rm{s^{-1}cm^{-2}}}$.}
\end{deluxetable}


\begin{table}[!bht]
\caption{Best Linear Fits of Line Luminosities and Stellar Properties.}
\label{linefits}
\begin{center}
\begin{tabular}{cccccccc}
 \tableline\tableline
                   Equation                                & \multicolumn{4}{c}{Best Fit~~~~} & \multicolumn{2}{c}{1:1 Fit\tablenotemark{w}} & $r$\tablenotemark{x}\\[+5pt]\cline{2-4}\cline{6-7}
                                                           &  $a$    &  $b$   & $\sigma$\tablenotemark{y} &  & a &  $\sigma$ & \\[+5pt]
\tableline
      $log(L_{\rm{C~IV}})={a}+{b}~log(L_{acc})$\tablenotemark{z}&-2.766  & 0.877 &0.60   &  & -2.609& 0.56 &  0.84\\[+5pt]
      $log(L_{\rm{FUV}})={a}+{b}~log(L_{\rm{C~IV}})$       & 1.059   & 1.023  &   0.19   &  &  0.960& 0.20 &  0.97\\[+5pt]
      $log(L_{\rm{FUV}})={a}+{b}~log(L_{acc})$             &-1.670   & 0.836  &   0.38   &  & -1.433& 0.47 &  0.85\\[+5pt]
\tableline
CTTS: $log(L_{\rm{C~IV}})={a}+{b}~log(L_{\rm{Ca~K}}) $     &-1.508   & 0.673  &   0.22   &  & -0.223& 0.57 &  0.78\\[+5pt]
WTTS: $log(L_{\rm{C~IV}})={a}+{b}~log(L_{\rm{Ca~K}}) $     & 0.609   & 1.232  &   0.19   &  & -0.509& 0.26 &  0.93\\[+5pt]
\tableline
CTTS: $log(L_{\rm{C~IV}})={a}+{b}~log(L_{\rm{H\alpha}})$   &-0.677   & 0.877  &   0.45   &  & -0.197& 0.41 &  0.77\\
\end{tabular}
\end{center}
\tablenotetext{w}{The 1:1 fits have the slope of the lines, $b$, fixed at 1.}
\tablenotetext{x}{$r$ is the Pearson linear correlation coefficient. The associated false-alarm probability is zero, 
         except that it is 0.38 for $log(L_{\rm{X}})$ and $log(L_{\rm{C~IV})}$.}
\tablenotetext{y}{Median scatter in dex in the data from the linear fit. }
\tablenotetext{z}{All the luminosities are in units of solar luminosity.}
\end{table}

\begin{table}
\caption{Luminosity in FUV Bandpasses$^a$}
\tabletypesize{\scriptsize}
\label{twhyaaumic}
\begin{tabular}{ccc}
\hline
Bandpass     &  TW Hya  &  AU Mic\\
\hline
\hline
$\log L_{phot}$         &  -0.77 &  -0.89\\
$L_{\rm{FUV}}^b$ (912--2000)  & -2.52 & -4.77 \\
$L_{\rm{FUV}}$ (1250--1700)  & -2.73 & -5.18\\
$L_{\rm Ca\ K}$ &  -4.18      & -5.39$^x$ \\
$L_{\rm H\beta}$ &  -3.33   & -- \\
$L_{\rm H\alpha}$ & -2.34 &  -4.56$^x$ \\
\hline
912--1000   &  -4.08 & -5.91       \\
~~\ion{C}{3} $\lambda977$ & -4.61 & -6.39 \\
1000--1100 & -3.95 &   -5.68       \\
~~\ion{O}{6}  $\lambda1032$ & -4.62 &-6.15  \\
1100--1200 & -3.89 &   -5.89        \\
~~\ion{C}{3} $\lambda1177$ & -4.49 & -6.23\\
1200--1300$^b$ &  -3.47 &  -5.80  \\
~~Ly$\alpha$ $\lambda1216$ & -1.95$^z$ & -4.49$^y$ \\
1300--1400  & -3.43 & -5.75  \\
~~\ion{C}{2}  $\lambda1335$ & -4.45 & -6.15  \\
1400--1500 &  -3.57 &  -6.19\\
1500--1600 & -3.21 & -5.71 \\
~~\ion{C}{4} $\lambda1549$ & -3.54 & -5.92 \\
1600--1700 & -3.31 & -5.72 \\
~~\ion{He}{2} $\lambda1640$ & -3.85 & -6.04  \\
1700--1800 & -3.31 & -5.86 \\
1800--1900 & -3.29 & -5.67 \\
1900--2000 & -3.37 & -5.89 \\
2000--3000 & -2.23 & -4.22 \\
~~\ion{Mg}{2} & -3.41 & -4.98 \\
\hline
\multicolumn{3}{l}{$^a$Luminosities in $\log L_{\lambda}/L_{bol}$; $^b$Excluding Ly$\alpha$;}\\
\multicolumn{3}{l}{$^x$\citet{Browning2010}; $^y$\citet{Wood2005};}\\
\multicolumn{3}{l}{$^z$\citet{Herczeg2004}.}\\
\end{tabular}
\end{table}

%
%
%

\clearpage

\clearpage

\clearpage

\clearpage

\clearpage

\clearpage

\clearpage

\clearpage

\clearpage

\clearpage

\clearpage

\clearpage


\clearpage

\clearpage

\clearpage

\begin{figure}[ht]
  \begin{center}
    \includegraphics[scale=0.75]{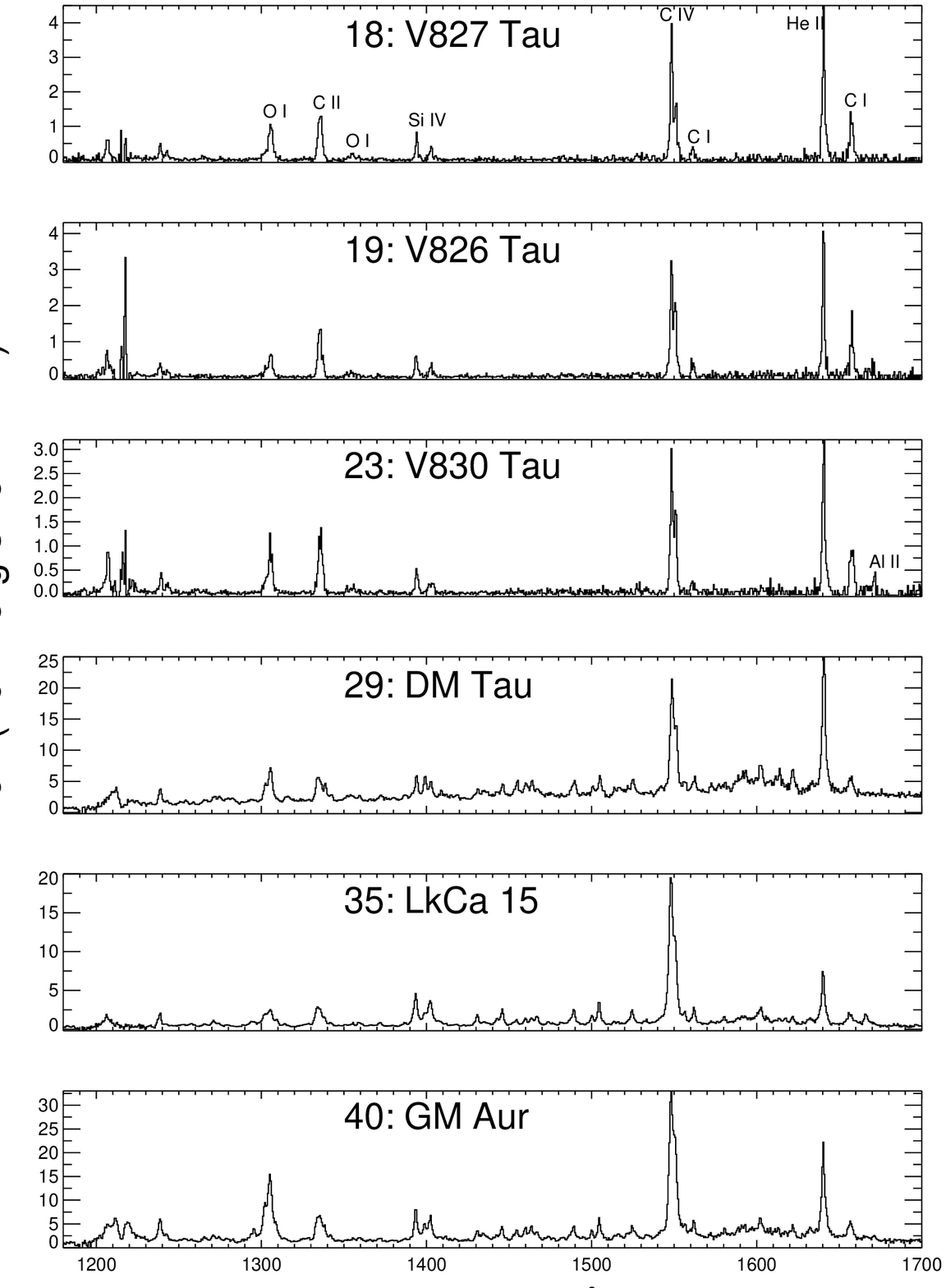}
       \caption{\emph{Online-only figure} STIS G140L plot 2.}
           \label{stisplot2}
              \end{center}
    \end{figure}

\begin{figure}[ht]
  \begin{center}
    \includegraphics[scale=0.75]{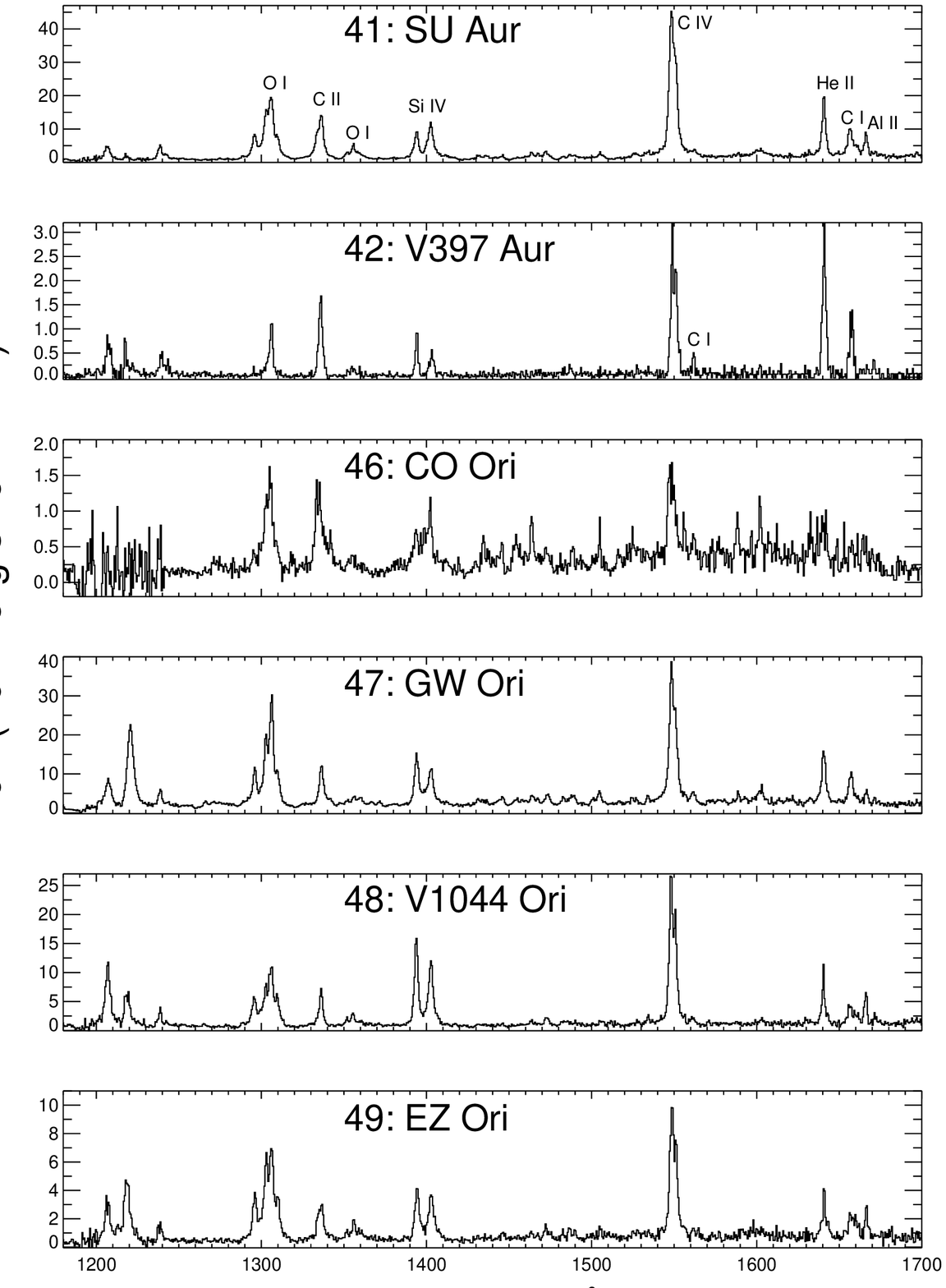}
       \caption{\emph{Online-only figure} STIS G140L plot 3.}
           \label{stisplot3}
              \end{center}
    \end{figure}

\begin{figure}[ht]
  \begin{center}
    \includegraphics[scale=0.75]{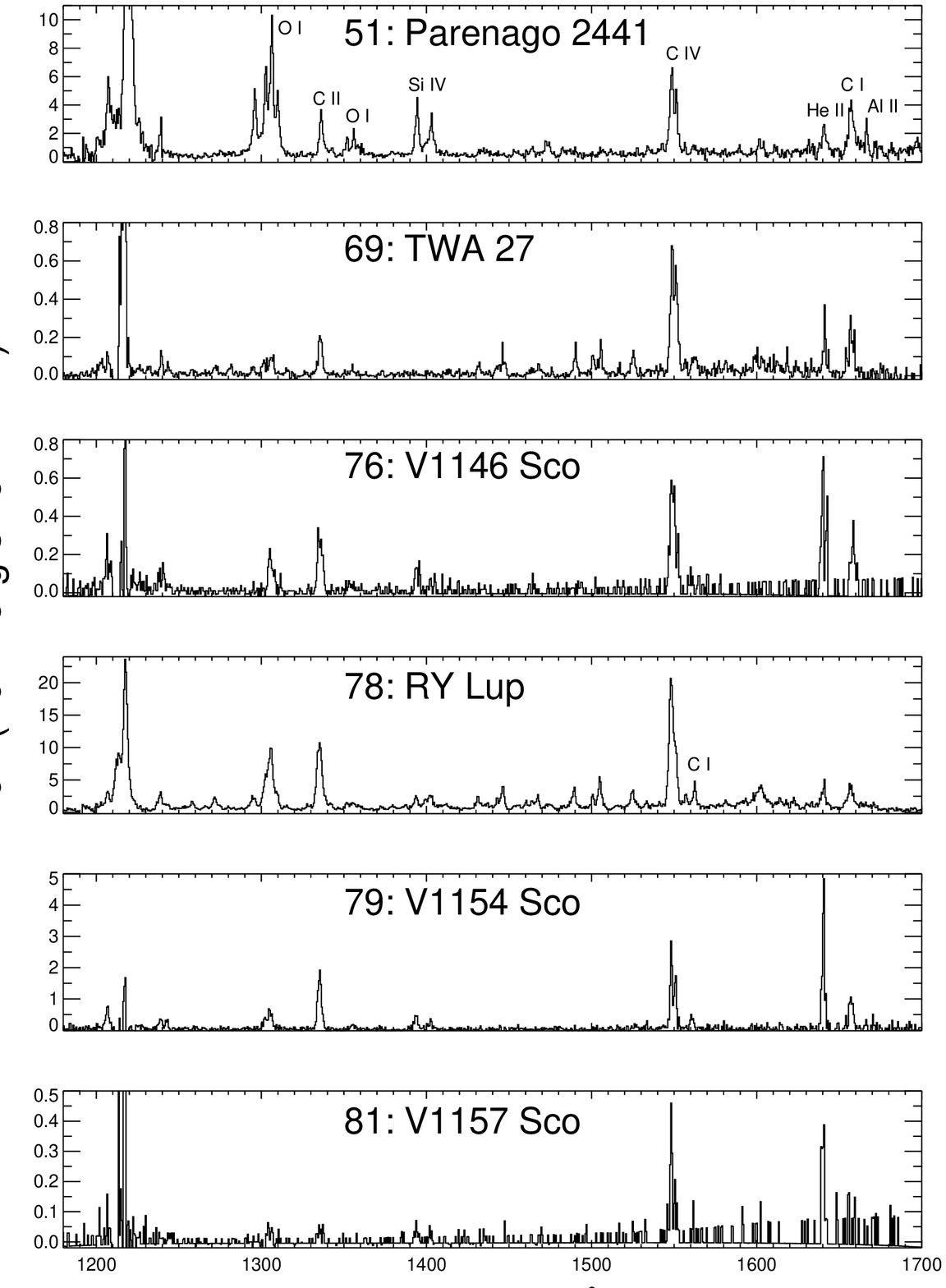}
       \caption{\emph{Online-only figure} STIS G140L plot 4.}
           \label{stisplot4}
              \end{center}
    \end{figure}

\begin{figure}[ht]
  \begin{center}
    \includegraphics[scale=0.75]{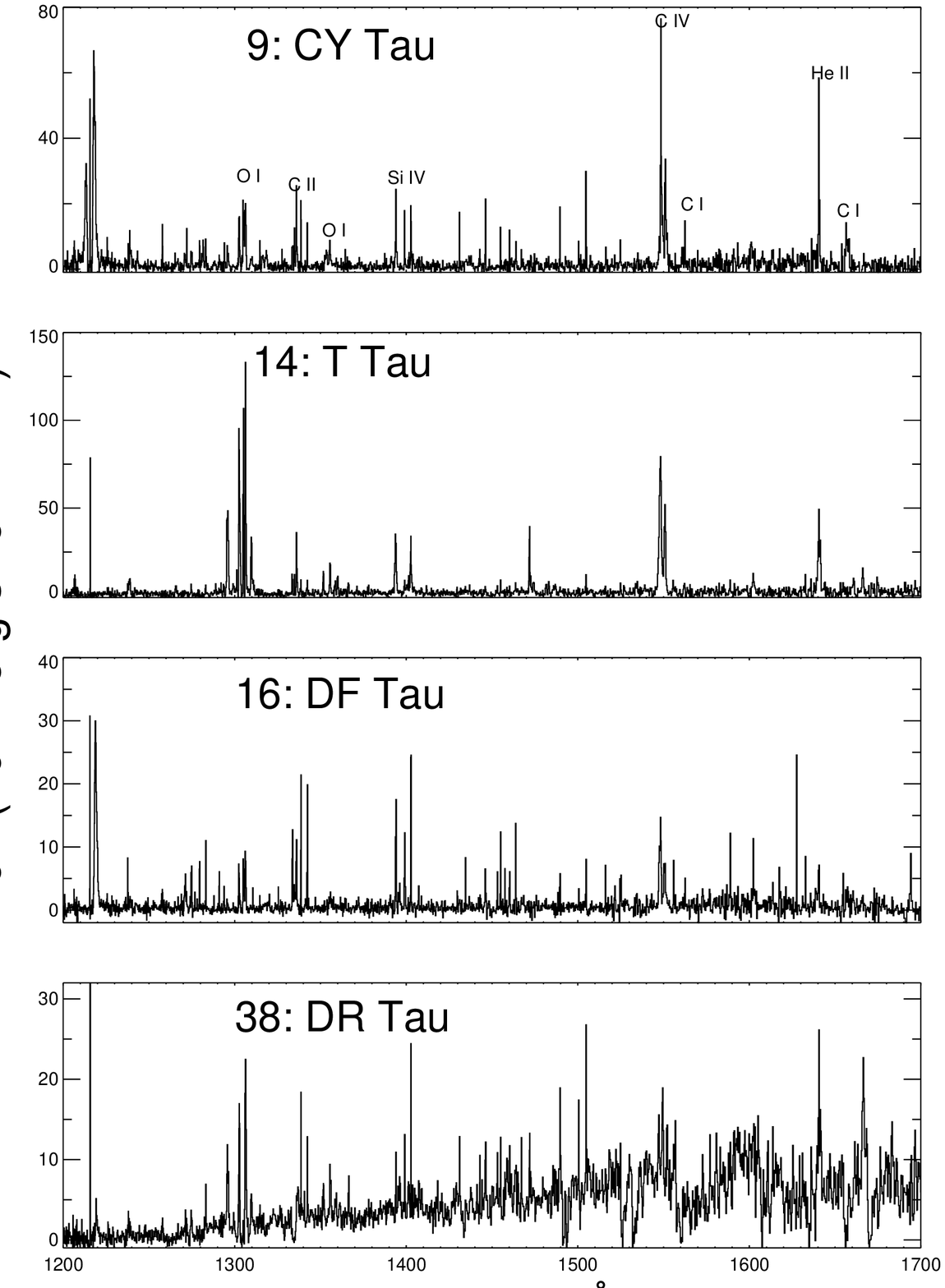}
       \caption{\emph{Online-only figure} STIS E140M plot 1.}
           \label{stise140mplot1}
              \end{center}
    \end{figure}

\begin{figure}[ht]
  \begin{center}
    \includegraphics[scale=0.75]{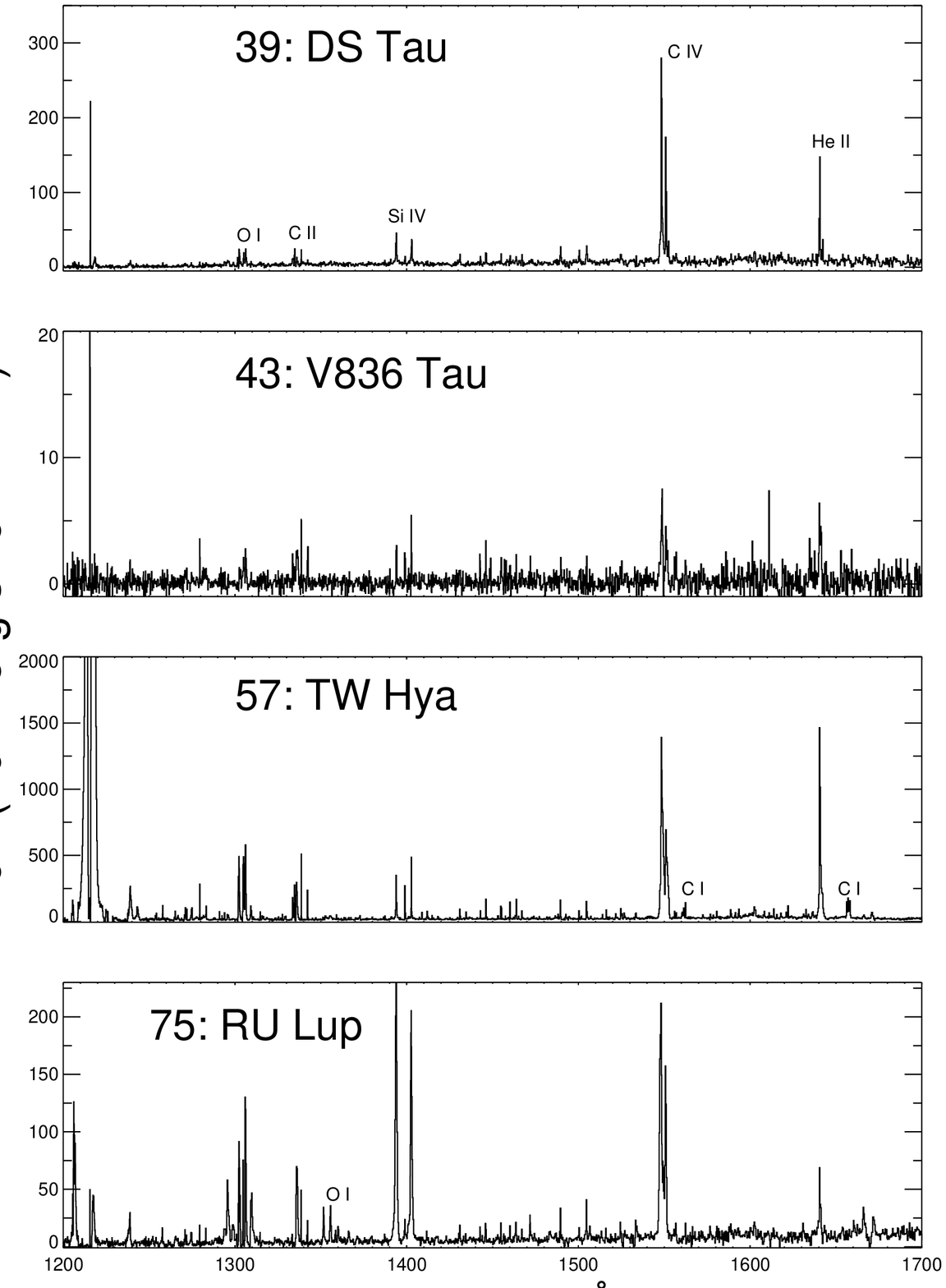}
       \caption{\emph{Online-only figure} STIS E140M plot 2.}
           \label{stise140mplot2}
              \end{center}
    \end{figure}

\begin{figure}[ht]
  \begin{center}
    \includegraphics[scale=0.75]{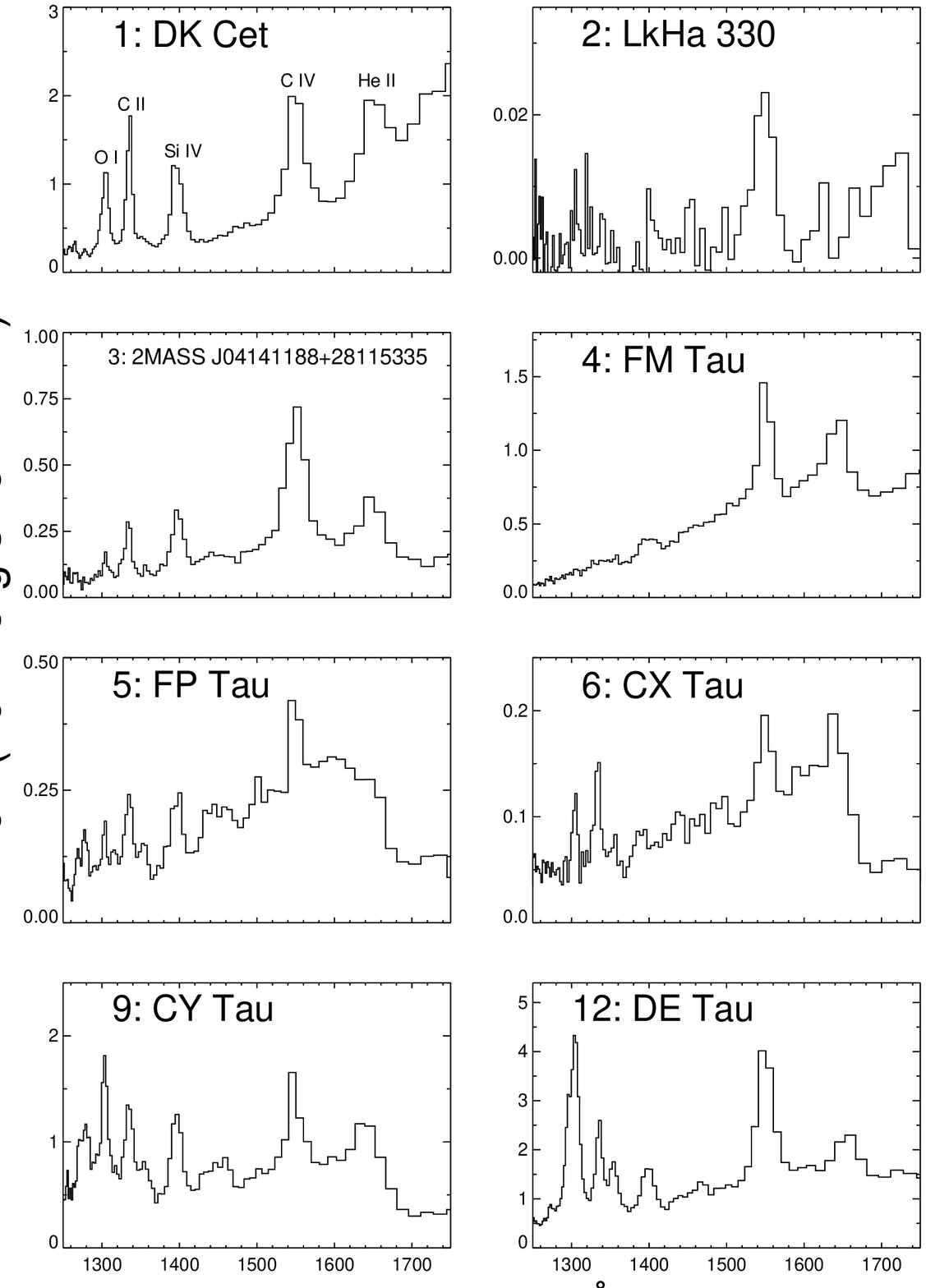}

       \caption{\emph{Online-only figure} ACS plot 1.}
           \label{acsplot1}
              \end{center}
    \end{figure}

\clearpage
\begin{figure}[ht]
  \begin{center}
    \includegraphics[scale=0.75]{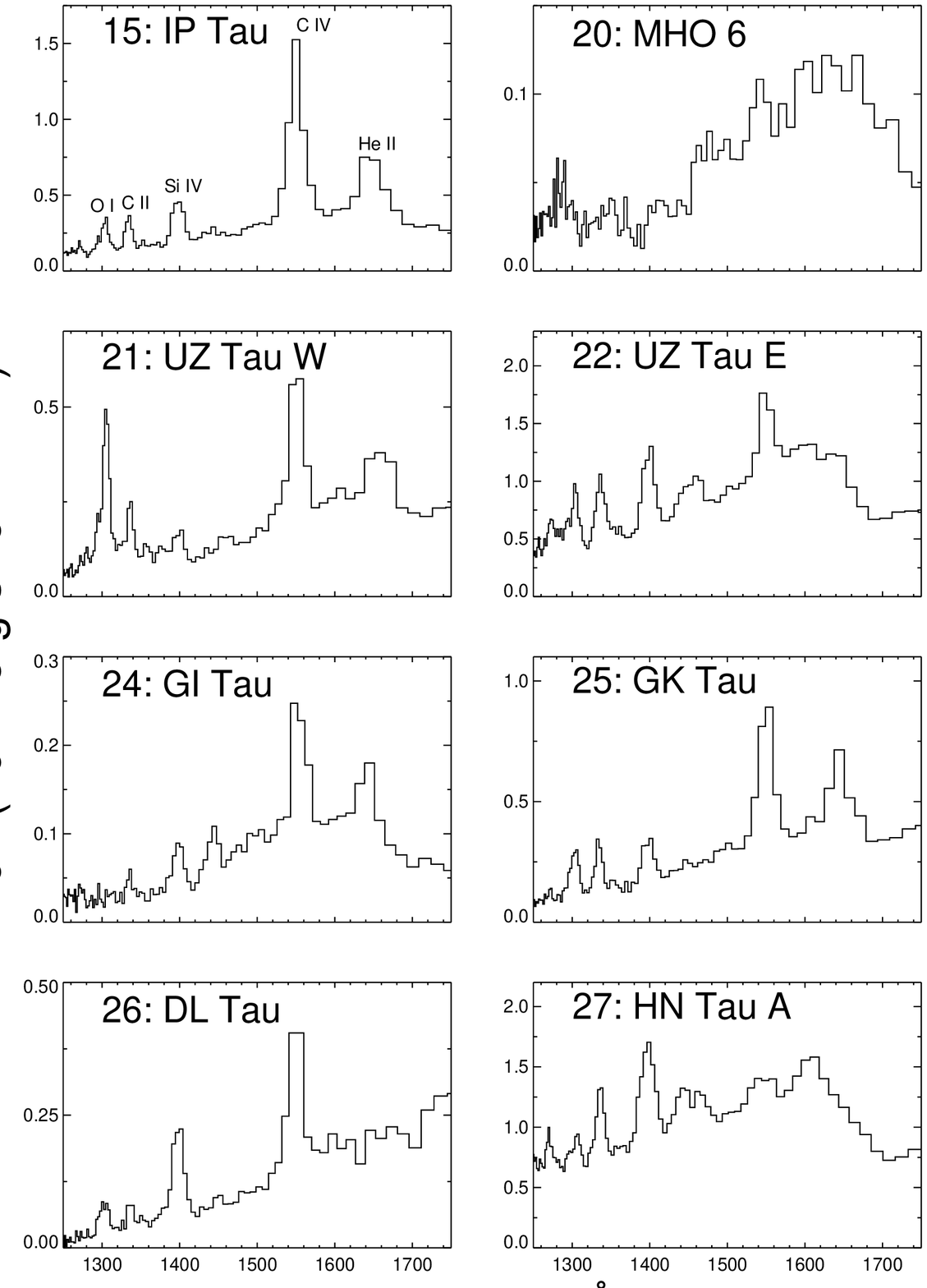}

       \caption{\emph{Online-only figure} ACS plot 2.}
           \label{acsplot2}
              \end{center}
    \end{figure}

\clearpage
\begin{figure}[ht]
  \begin{center}
    \includegraphics[scale=0.75]{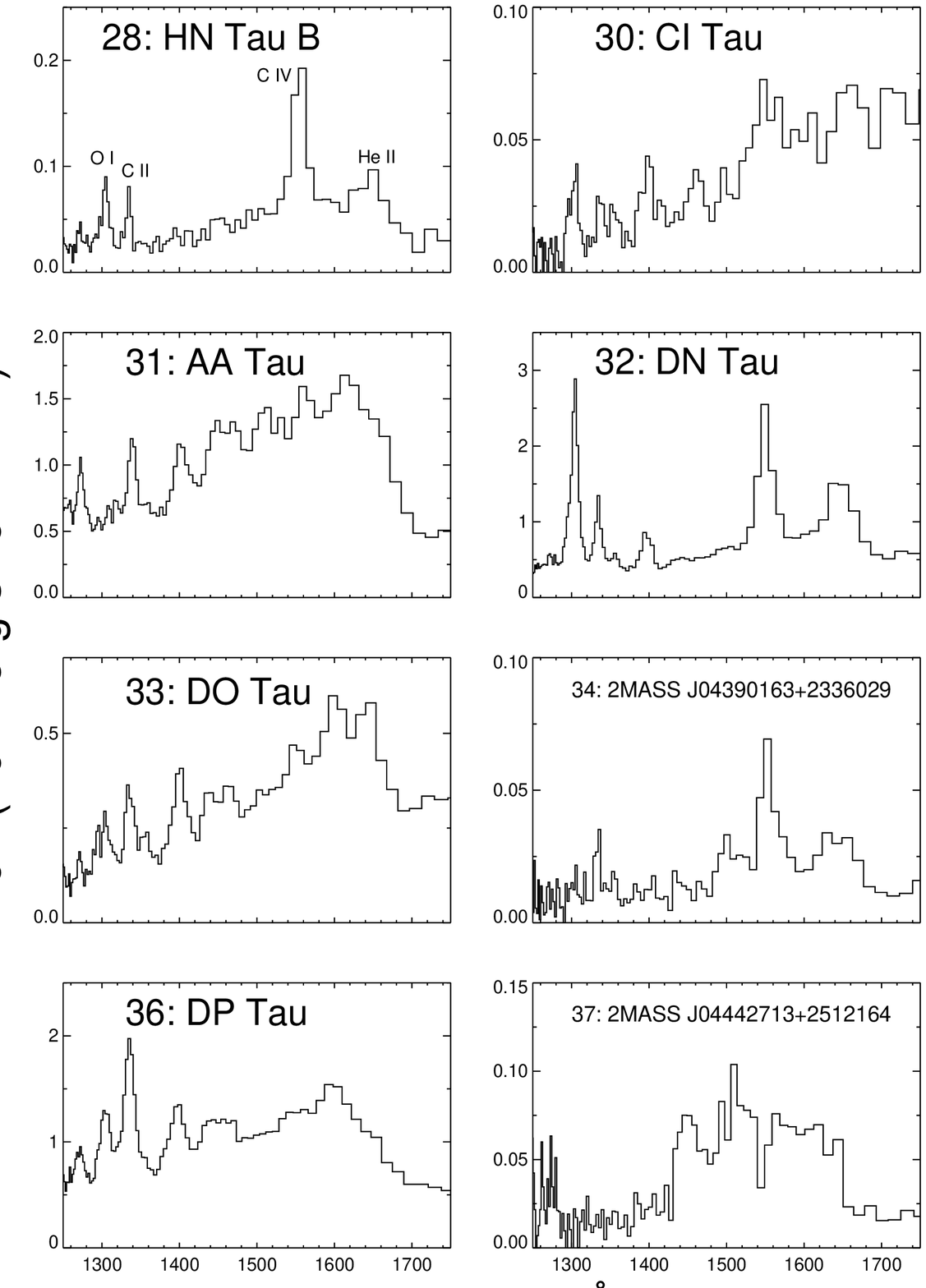}

       \caption{\emph{Online-only figure} ACS plot 3.}
           \label{acsplot3}
              \end{center}
    \end{figure}

\clearpage
\begin{figure}[ht]
  \begin{center}
    \includegraphics[scale=0.75]{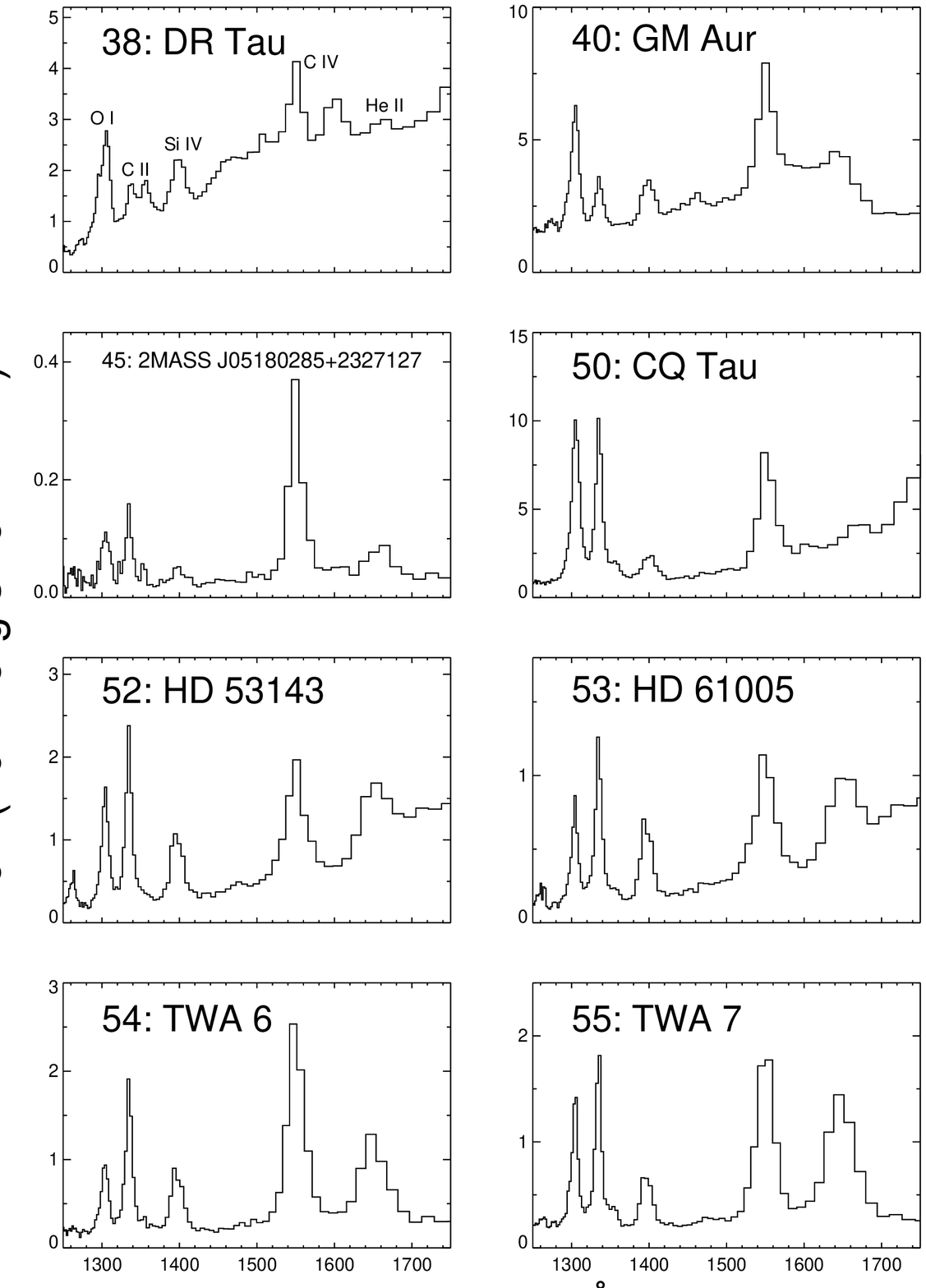}

       \caption{\emph{Online-only figure} ACS plot 4.}
           \label{acsplot4}
              \end{center}
    \end{figure}

\clearpage
\begin{figure}[ht]
  \begin{center}
    \includegraphics[scale=0.75]{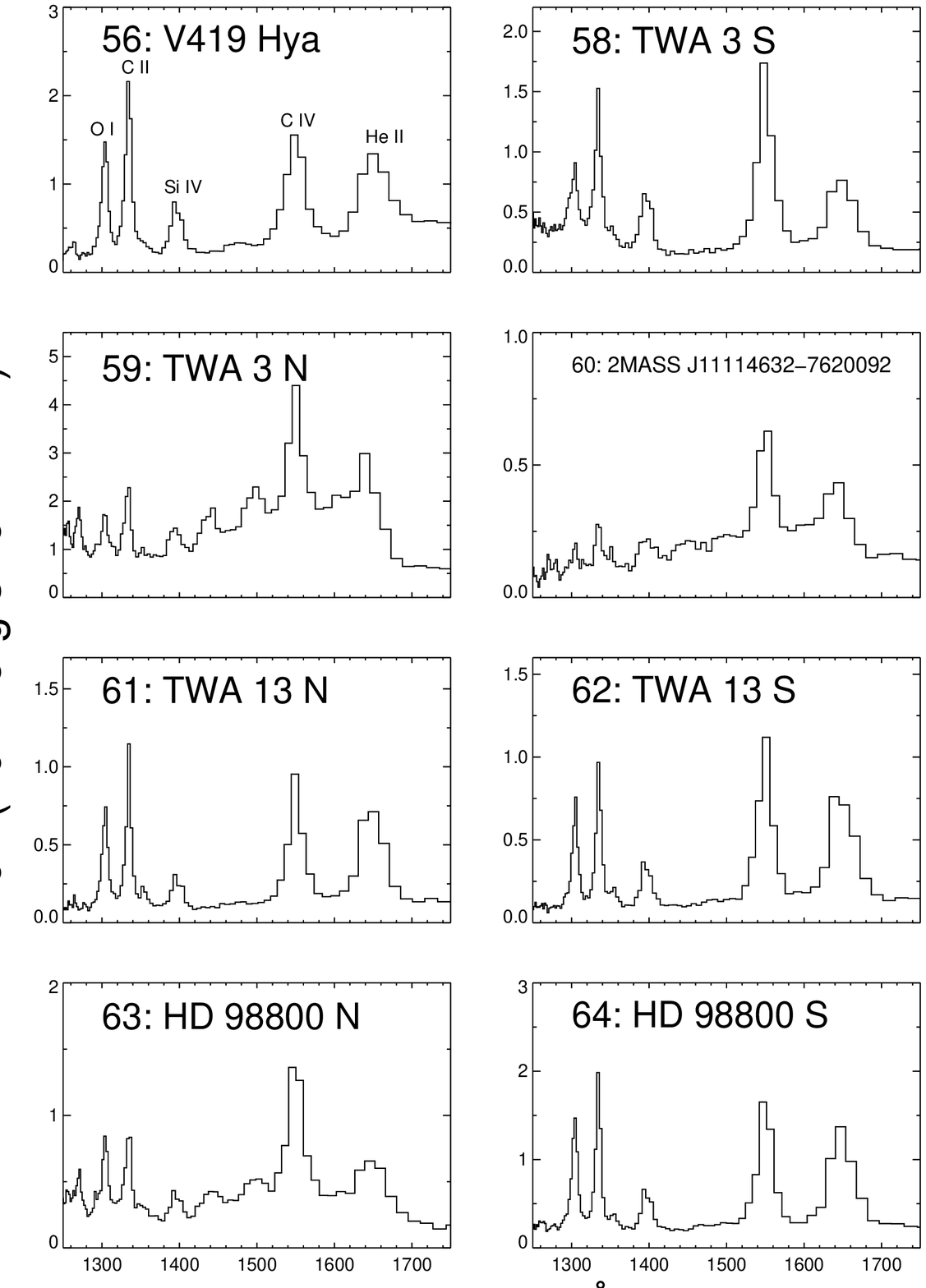}

       \caption{\emph{Online-only figure} ACS plot 5.}
           \label{acsplot5}
              \end{center}
    \end{figure}

\clearpage
\begin{figure}[ht]
  \begin{center}
    \includegraphics[scale=0.75]{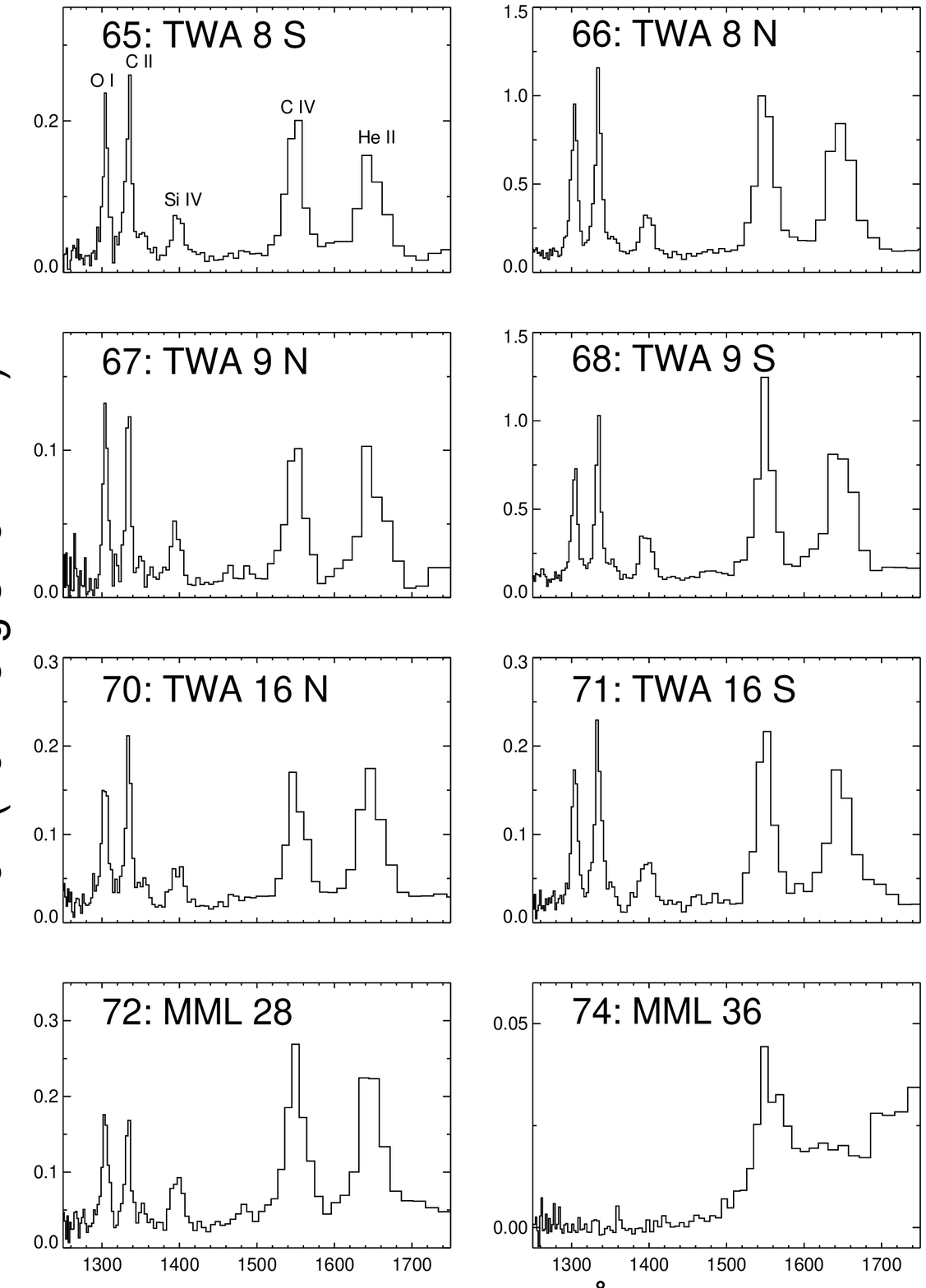}
       \caption{\emph{Online-only figure} ACS plot 6.}
           \label{acsplot6}
              \end{center}
    \end{figure}

\clearpage
\begin{figure}[ht]
  \begin{center}
    \includegraphics[scale=0.75]{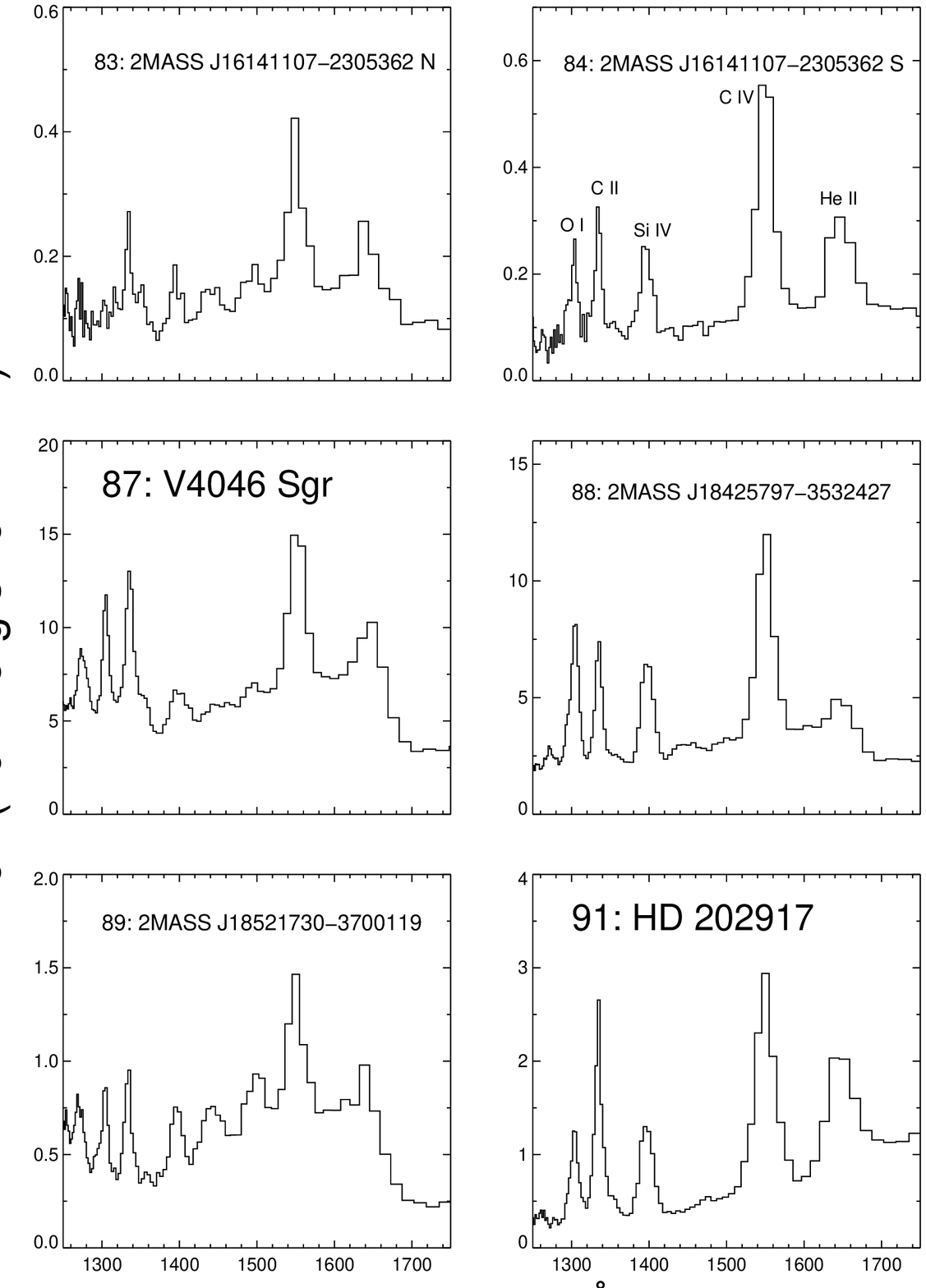}
       \caption{\emph{Online-only figure} ACS plot 7.}
           \label{acsplot7}
              \end{center}
    \end{figure}

\clearpage
\begin{figure}[ht]
  \begin{center}
    \includegraphics[scale=0.75]{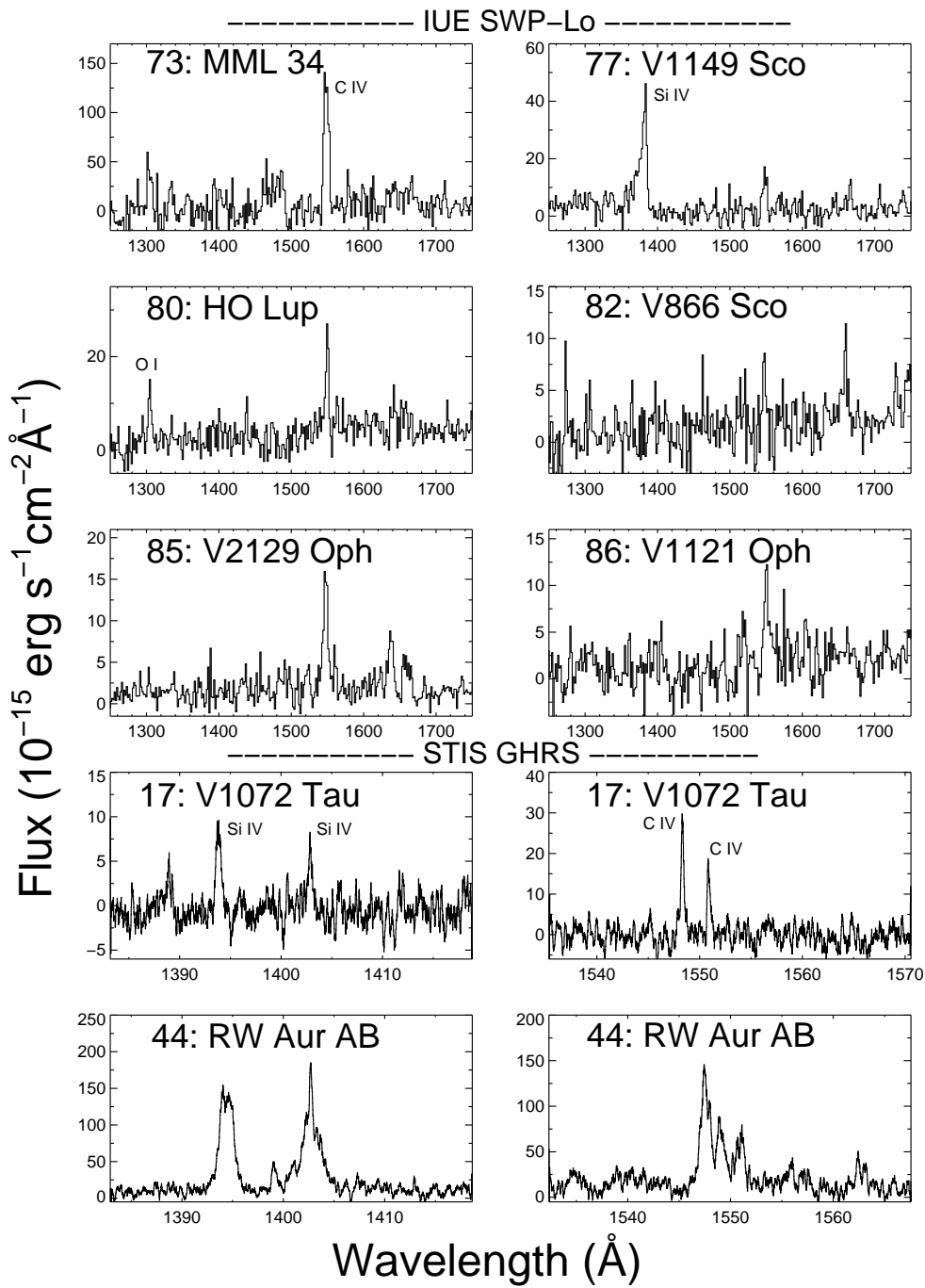}
       \caption{\emph{Online-only figure} IUE and GHRS plot.}
           \label{IUEplot}
              \end{center}
    \end{figure}

\end{document}